\documentclass[twocolumn, pra, aps, superscriptaddress,longbibliography,showpacs,amsmath,amssymb,floatfix]{revtex4-1}                   
\usepackage{graphics}
\usepackage{amssymb}
\usepackage{amsmath}
\usepackage{epsfig}
\usepackage{color}
\usepackage{tabu}
\usepackage{mathtools}
\usepackage[colorlinks,linkcolor=blue,anchorcolor=blue,citecolor=blue,urlcolor=blue]{hyperref}
\usepackage{physics}
\usepackage{float}
\usepackage{diagbox}
\usepackage{verbatim}

\usepackage{xcolor}

\begin{document}

\title {Landau-Zener-St\"uckelberg interference in edge state pumping}
\author{Y. Liu}
\email{yangjie@hubu.edu.cn}
\affiliation{Department of Physics, School of Physics, Hubei University, Wuhan 430062, China}
\affiliation{Department of Optics \& Optical Engineering, School of Physical Sciences, University of Science and Technology of China, Hefei 230026, China}
\author{Xiaoshui Lin} 
\affiliation{CAS Key Laboratory of Quantum Information, University of Science and Technology of China, Hefei 230026,  China}
\affiliation{Synergetic Innovation Center of Quantum Information and Quantum Physics, University of Science and Technology of China, Hefei 230026, China}
\author{Ming Gong}
\email{gongm@ustc.edu.cn}
\affiliation{CAS Key Laboratory of Quantum Information, University of Science and Technology of China, Hefei 230026, China}
\affiliation{Synergetic Innovation Center of Quantum Information and Quantum Physics, University of Science and Technology of China, Hefei 230026, China}
\affiliation{Hefei National Laboratory, University of Science and Technology of China, Hefei 230088, China}
             
\begin{abstract}
The adiabatic edge state pumping (ESP) in one dimensional model, which has important applications in topological phase transition and quantum simulation, has been widely performed in both theories and experiments. This phenomenon has been observed in some systems with sizes $L = 9 - 100$, and it seems that due to the topological protection, the ESP can be survived even in the presence of weak random potential. Yet the fundamental issues of adiabaticity for this process have not been clarified. In this paper, we revisit this problem and show that this process involves two non-adiabatic points during the transition between the edge state and bulk state, yielding non-abadiatic physics. As a result, the ESP can be described by the Landau-Zener-St\"{u}ckelberg (LZS) interference process, in which the relative phase between the edge state and the bulk state determine the fate of the edge state during pumping. Furthermore, in a relatively long chain with weak disorder, the ESP can break down due to the anti-crossing of the edge state and the bulk edge states. We unveil these physics in terms of non-adiabaticity. The new mechanisms for ESP unveiled in this work is readily accessible in experiment, and shall therefore offer a down-to-earth platform for the 
intriguing LZS dynamics in terms of edge states. 
\end{abstract}

\maketitle

{\it Introduction}: 
The discovery of topological matters have stimulated widespread investigations of their edge states ensured by bulk-edge correspondence in quantum simulation \cite{Qi2011, Hasan2010, Schnyder2008Classification,Ozawa2019a, Georgescu2014Quantum}. In one dimensional (1D) topological models, the edge states are localized at the open ends, which can gradually transfer from one end to the other by adiabatically tuning the parameters, termed as Thouless (or adiabatic) pumping \cite{Thouless1983Quantization, Citro2023, Aleiner1998Adiabatic, Nakajima2016Topo, SunY2022Non-Abelian, Jurgensen2021, Jurgensen2023, Cerjan2020}. It has been believed that when the pumping process is slow enough, this kind of adiabatic process can always be achieved. Following this tenet, the adiabatic pumping in a chain with quasiperiodic potentials were experimentally investigated using coupled single-mode waveguides \cite{Lahini2009, Kraus2012, Rechtsman2013, Tambasco2018Quantum, WangY2019, Jurgensen2021}. This idea was also generalized to higher dimensions in terms of synthetic dimensions \cite{YuanL2018, Zilberberg2018, YangZ2020}. Since the edge states are protected by topology, this adiabatic pumping withstands weak disorders \cite{Nakajima2021, Kolodrubet2018Floquet-Thouless} and nonlinear interactions \cite{Jurgensen2021, Tangpanitanon2016, Mostaan2022, Jurgensen2023, Qian2011Quantum}. In recent years, this adiabatic pumping has been widely examined using electromechanical patches \cite{XiaY2021, WangS2023, Rosa2019, Frank2022Boosting}, acoustic waveguides \cite{Apigo2019, NiX2019Observation, ChenZ2021, You2022Observation, HuP2024Hearing}, photonic crystals \cite{KeY2016, Meier2016, Alpeggiani2019Topo, ChengQ2022, Adiyatullin2023}, and even in the Floquet-Bloch band \cite{Upreti2020, Minguzzi2022, Adiyatullin2023}. The chain sizes for all systems above range about 9 - 100. The related adiabatic transport of matter in terms of Thouless pumping even with many-body interaction was also explored in experiments \cite{Schweizer2016, Viebahn2024Interactions, You2022Observation} for applications in quantum matter transport, state transfer and information processing. 

This edge state pumping (ESP) is fundamentally different from the adiabatic transition in the two-level system \cite{Cohen2019Geometric, Berry1984Geometry}, in which the adiabaticity is guaranteed by the finite gap between the two levels. In the continuum limit for many-level system, the number of energy levels is mainly determined by the lattice size $L$, in which the larger the size $L$ is, the smaller the level spacing $\Delta$ will be. Thus there are two competitive ingredients to influence the adiabaticity during pumping: the overlap between the edge state and its nearest neighboring bulk state, and the energy gap between them, both of which decrease with the increasing of $L$. Therefore when the former term decreases faster than the latter, the adiabatic condition can still be fulfilled; otherwise, it becomes possible that the adiabatic condition may break down, yielding non-adiabatic evolution. Therefore, while the ESP has been verified in experiment \cite{Kraus2012, Rechtsman2013, Switkes1999Adiabatic, Lahini2009, Rosa2019, Frank2022Boosting, Verbin2015Topological, Verbin2013, Zilberberg2018, YangZ2020, Nakajima2021, Kolodrubet2018Floquet-Thouless, Jurgensen2021, Tangpanitanon2016, Mostaan2022, You2022Observation, Jurgensen2023, XiaY2021, WangS2023, Apigo2019, NiX2019Observation, ChenZ2021, HuP2024Hearing,KeY2016, Meier2016, Alpeggiani2019Topo, ChengQ2022, Adiyatullin2023, Viebahn2024Interactions, XiangZ2023Simulating, Rajagopal2019, Cerjan2020}, the stringent validity of adiabatic condition, which is essential for ESP, remains yet to be verified. 
Furthermore, the disorder effect in ESP also needs to be clarified \cite{Rajagopal2019, YuS2020, Cerjan2020, Tangpanitanon2016}. These two issues will affect ESP severely in the large $L$ limit with vanished coupling between the edge states \cite{ChenZ2021}. 

We then revisit this problem and present a new picture for the ESP in these models. (I) We show that there are two non-adiabatic points (NAPs) during the ESP process, at which the adiabaticity breaks down even in the slowly-varying limit. At the doublet of NAPs, the ESP process reduces to the Landau-Zener-St\"{u}ckelberg (LZS) interference 
\cite{Berry2009Transitionless, Shevchenko2010LZS, WangL2015Atom-interferometric, Boyers2020, ChenZ2021, QiaoX2023Nonlinear}, resulting in transient oscillation of pumping efficiency as evolution time $T$ changes. (II) In the presence of weak disorder when the localization length $\xi$ is larger than the system size $L$, the anticrossing of the bulk states and edge state breaks adiabaticity when the edge state is immersed into the bulk states \cite{Lahini2009}. In this specific case, the ESP will cease to happen. These new mechanisms can be readily examined in simulating platforms with ultracold atoms \cite{Schomerus2013Topo, Viebahn2019, Nakajima2021}, optical and acoustic waveguides \cite{XiaY2021, WangS2023}
and electric circuits \cite{Li2023Observation, Mei2018Robust}, etc.. 

\begin{figure}[hbtp!]
  \includegraphics[width=0.45\textwidth]{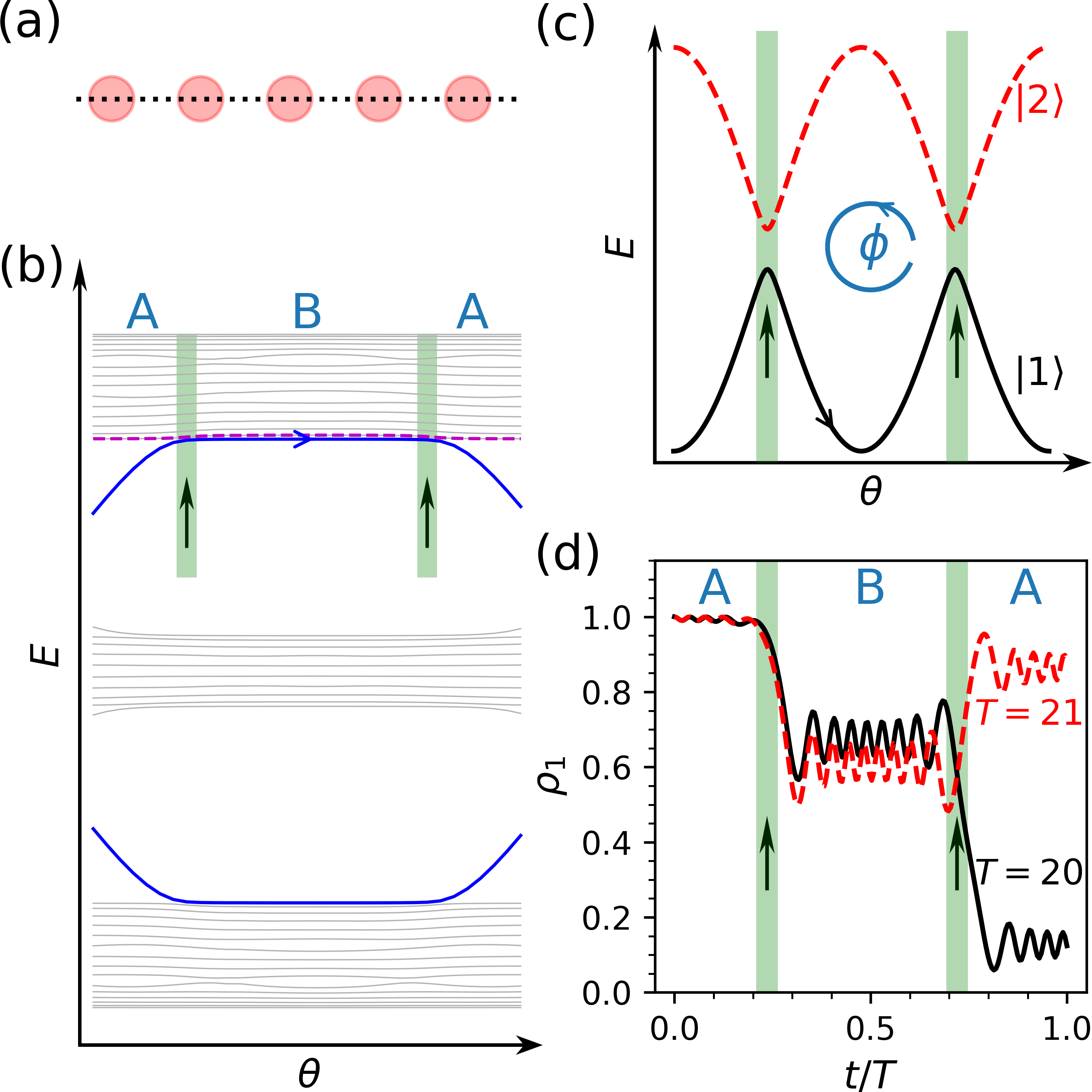}
  \caption{(a) The 1D tight-binding model considered in this work. (b) Energy levels of the edge state (in solid blue lines) and bulk state (in solid gray lines). (c) The two-level LZS model with relative phase $\phi$ accumulated during evolution, being the phase between the two levels. (d) Occupation probability $\rho$ of the first state $\vert \psi_1\rangle$ in the LZS interference model (c). Parameters for the LZS model are $A=4$, $g=0.4$ and $T = 20$ (black solid line) 
  and 21 (red dashed line). }
  \label{fig-fig1}
\end{figure}

{\it Physical model and LZS interference}:  We adopt the off-diagonal Andre-Aubry-Harper (AAH) model \cite{Aubry1980Analyticity, Ganeshan2013Topological, Yahyavi2019Generalized, Liu2018Mobility} as a testbed to present these new mechanisms 
\begin{align}
\label{eq-Hamiltonian}
\hat{H}=\sum_n V_n \hat{c}_n^\dagger \hat{c}_{n+1}+ \text{h.c.} + W\xi_n \hat{c}_n^\dagger \hat{c}_n, 
\end{align}
with incommensurate potential $V_n = V[1 + \lambda\cos(\pi \alpha n + \theta)]$, 
where $V$ is the hopping strength between neighboring sites, $\alpha$ is the irrational constant and $\hat{c}_n$ ($\hat{c}_n^\dagger$) is the annihilation (creation) operator at site $n$ [see Fig. \ref{fig-fig1} (a)]. Hereafter, we use $V = 8/15$, $\lambda = 0.6$, $\alpha = (\sqrt{5} +1)/2$ (parameters from Ref. \cite{Kraus2012}), with uniform disorder $\xi_n \in [-1/2, 1/2]$. The adiabatic parameter $\theta$ calibrates the process when the edge state transfers from the left end to the right within a finite pump time $T$, and the energy levels in this model versus the phase $\theta$ are presented in Fig. \ref{fig-fig1} (b), in which the two thick solid blue lines represent the edge modes, respectively, with opposite energies. We focus on the pumping of edge state with $E > 0$ in this paper. To this end, we expand the wave function using $\vert\psi (t) \rangle =\sum_n c_n(t)\vert n\rangle$, where $|n\rangle = \hat{c}_n^\dagger|0\rangle$, 
and the coefficient $c_n$ is determined by Schr\"odinger equation $i\partial_t\vert \psi(t)\rangle = \hat{H}(t)\vert\psi(t)\rangle$, with $\theta = \theta(t)$ to be the time-dependent tuning parameter.  

\begin{figure}[hbtp!]
  \includegraphics[width=0.45\textwidth]{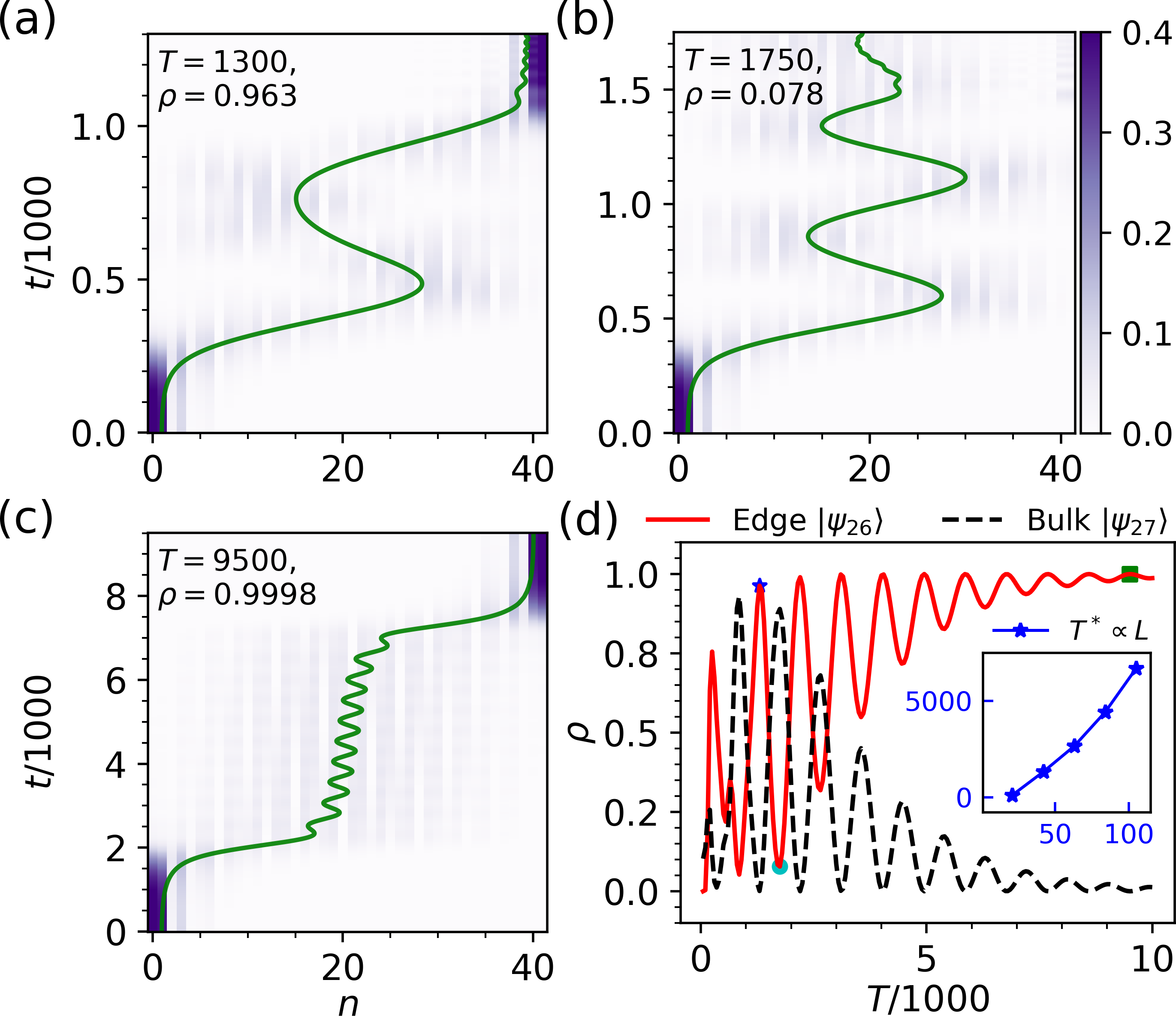}

  \caption{Distribution of wave function density $|\psi(x_i, t)|^2$ in spacetime and its mean trajectories $\bar X(t)= \sum_i x_i \vert\psi(x_i, t)\vert^2$ (green lines) for three pump times: (a) $T=1300$, (b) $T=1750$ and (c) $T=9500$,  with $L = 42$ and without disorder. 
  (d) Occupation probability $\rho_n$ versus pump time $T$ for edge state $\vert \psi_{26}\rangle$ and bulk state $\vert \psi_{27}\rangle$. Note that the three pump times in panels (a-c) are chosen from three extrema on $\rho_{26}$ marked in (d). Inset shows the optimum pump time $T^*$ for the first maximum [blue asterisk in (d)] versus the chain length $L$. }
    \label{fig-fig2} 
\end{figure}

Our central idea is that during the ESP, there are two points at which the adiabatic condition is most vulnerable to break. This can be understood as follows. When the edge state is fully localized at the end, the overlap induced by variation of Hamiltonian $\delta H$, which may be defined as $O \sim {\vert \langle \psi_{\rm e}(\theta)\vert \delta H \vert \psi_{\rm b}(\theta)\rangle\vert}$ with $\vert\psi_{\rm e}\rangle$ for the edge state and $\vert\psi_{\rm b}\rangle$ the bulk state respectively, will decay as $1/L$. Meanwhile, the energy difference $\Delta$ between the bulk and edge states will approach a constant; see regime A in Fig. \ref{fig-fig1} (b).  Furthermore, for $\theta$ in regime B in Fig. \ref{fig-fig1} (b) when the edge states become extended, the overlap between them goes $O \sim 1/L$, while the energy gap between these two states also scales as $\Delta \sim 1/L$. Under these two conditions the the adiabatic condition can be easily fulfilled by controlling the sweeping rate even in the large $L$ limit \cite{Grabarits2024Floquet-Anderson}. The adiabaticity may break down between the regimes A and B with the increasing of system size $L$, which are marked by arrows as NAPs. 

Thus when the wave functions is restricted to occur between the edge state and the nearest neighboring bulk state and by projecting the wave function to these two states, the dynamics will be reduced to the LZS interference. To gain some insight into this process, let us first consider the following two-level model [Fig. \ref{fig-fig1} (c)]
\begin{equation}
\hat{H} = g \hat\sigma_x + A\hat\sigma_z\cos(\frac{2\pi t}{T}),
\label{eq-twolevel}
\end{equation} 
where $\hat\sigma_x$ and $\hat\sigma_z$ are Pauli matrices, $g$, $A$ are tunneling amplitude and bias amplitude respectively, and $T$ is the total evolution time. The Landau-Zener tunneling happens at $t = T/4$ and $T = 3T/4$. In Fig. \ref{fig-fig1} (d) the dynamics of wave functions is presented, showing that by slightly changing $T$, the final state can be significantly different. The interference depends strongly on the relative phase $\phi$ between the two levels [Fig. \ref{fig-fig1} (c)], which has been widely verified in experiments \cite{Huang2011LZS, Quintana2013Cavity, Korkusinski2017LZS, WangS2023Photonic}. 

{\it ESP without disorder}: We first consider the case without disorder ($W = 0$). Initially, we assume the wave function as the edge state $|\psi_n(\theta)\rangle$ and define occupation probability 
\begin{equation}
\rho_n = |\langle \psi(t)| \psi_n(\theta)\rangle |^2,
\end{equation}
where $|\psi_n(\theta)\rangle$ is the $n$-th instantaneous eigenstate with phase $\theta(t)$ and $|\psi(t)\rangle$ is the time-dependent wave function. The calculation results for different $T$ are presented in Fig. \ref{fig-fig2} (a) - (c) for size $L = 42$, and $n =26$ as the index of the edge state. For $T = 1300$ and $9500$, the left-end state successfully transfers to the right end. However, for $T = 1750$, the state will \emph{not} evolve to the right end and do so for $T=9500$ in panel (d). These results are in stark contrast to adiabatic theory in which the larger $T$ is, the better the adiabaticity will be \cite{Berry1984Geometry}. In Fig. \ref{fig-fig2} (d), the oscillation of $\rho_{26}$ and $\rho_{27}$ versus pump time $T$, is a  
signature of strong nonadiabaticity before entering the adiabatic limit. Furthermore, our result means $\rho_{26} + \rho_{27} \simeq 1$, indicating that only these two states are important during evolution, which justifies the reduced LZS dynamics of two levels. This oscillation was found numerically in Ref. \cite{Longhi2019Topo}, but with its fundamental origin not revealed. The back and forth of the mean trajectories $\bar{X}(t)$ can be readily measured in the currently available experiments. In the inset of (d), we plot the the pump time $T^*$ for the first maximum ESP (in blue asterisk) versus system length $L$, showing that $T^* \propto L$, which indicates that the longer the chain is, the longer time $T^*$ it takes to achieve efficient ESP. 

\begin{figure}[hbtp!]
  \includegraphics[width=0.45\textwidth]{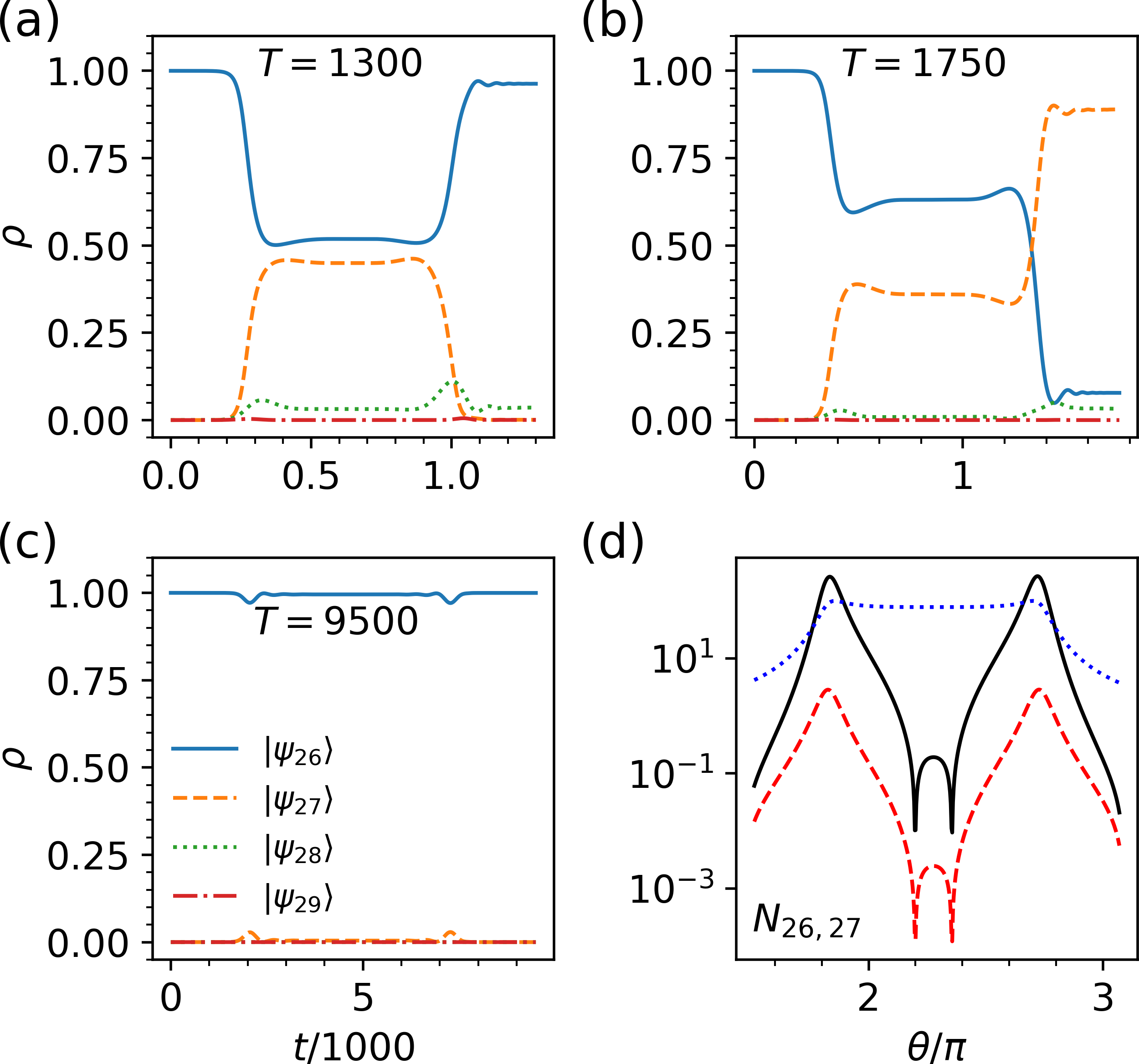}
  \caption{Evolution of $\rho(t)$ for the edge state $\vert \psi_{26}\rangle$ and the three nearest bulk states $\vert \psi_{27}\rangle$, $\vert \psi_{28}\rangle$ and $\vert \psi_{29}\rangle$ for pump time (a) $T=1300$, (b) $T=1750$ and (c) $T=9500$ (Here, $L = 42$). (d) Non-adiabaticity ${N}_{26, 27}$ (solid line), overlapping $O_{26, 27}$ (dashed line) and reciprocal gap $\vert \Delta_{26, 27}\vert^{-1}$ (dotted line) as a function of the phase $\theta$.  } 
    \label{fig-fig3}
\end{figure}

To clarify the LZS dynamics, we plot the evolution of $\rho_n$ ($n = 26 - 29$) for $L=42$ for three typical $T$ in Fig. \ref{fig-fig3} (a) - (c). A common feature is that the tunnelling occurs only at the two NAPs. In panel (a) at the first NAP, $\rho_{26}$ drops suddenly to about 0.5, lending the other half portion to the bulk $\rho_{27}$. However at the second NAP, the edge state $\vert\psi_{26}\rangle$ will then reclaim the occupation from the bulk $\vert\psi_{27}\rangle$, reaching the unity probability. This is a surprising example to realise a perfect ESP yet with \emph{non-adiabatic} process. In (b) with an even-larger pump time $T=1750$, the edge state finally transfers to the bulk after two tunnellings at the NAPs, with $\rho_{26} \sim 0$. In (c) when $T$ is large enough, the ESP is achieved in the adiabatic limit. This dependence of occupation probabilities $\rho_n$ with pump time $T$ corroborates the LZS process shown in Fig. \ref{fig-fig1} (d). 

In order to determine the precise positions of the NAPs for the breaking of adiabaticity, we consider the evolution of quantum state as \cite{Berry1984Geometry}
\begin{equation}\label{eq-expansion}
\vert \psi(t)\rangle=\sum_{m}c_m(t) \exp(-i\int_{0}^t {\rm d}t'E_m(t')) \vert \psi_m (t) \rangle,
\end{equation}
where $E_m$ is the eigenenergy for instantaneous eigenvector $\vert \psi_m (t)\rangle$. Substituting Eq. \eqref{eq-expansion} into Schr\"odinger equation yields
\begin{eqnarray}
\partial_{t}c_n(t)&=&-c_n(t)\langle\psi_n\vert\partial_t\vert\psi_n\rangle-
\sum_{m\neq n}c_m\langle\psi_n\vert\partial_t\vert \psi_m\rangle\cdot
\nonumber\\&&
\exp(-i\int_{0}^t {\rm d}t'[E_n(t')-E_m(t')]).
\end{eqnarray}
Then the adiabatic condition requires that for $\forall {m\neq n}$, 
$\quad \vert \langle\psi_n\vert\partial_t\vert\psi_m\rangle \vert \ll \vert \langle\psi_n\vert\partial_t\vert \psi_n\rangle \vert$, and from Schr\"odinger equation a sufficient condition for adiabaticity can be \cite{Aharonov1987Phase, Amin2009, Marzlin2004Inconsitency, Riva2021Adiabatic}  
\begin{equation}
\vert \langle\psi_n\vert\partial_t\vert\psi_m\rangle \vert=\frac{{\rm d}\theta}{{\rm d}t} \vert \langle\psi_n\vert\partial_\theta\vert\psi_m\rangle \vert \ll \vert E_n-E_m\vert. 
\end{equation}
With this we define the non-adiabaticity $\mathcal{N}_n$ for $\vert\psi_n(\theta)\rangle$, irrespective of the sweeping rate ${\rm d}\theta/{\rm d}t$, as 
\begin{eqnarray}
\label{eq-An}
\mathcal{N}_n(\theta)=\sum_{m\neq n} N_{n,m} (\theta)
\end{eqnarray}
where $N_{n,m}={O_{n,m}}/{\Delta_{n, m}}, O_{n,m}={\vert \langle \psi_n(\theta)\vert \partial_\theta \vert \psi_{m}(\theta)\rangle\vert}$ and $\Delta_{n, m}={\vert E_n- E_m\vert}$. The larger this value $\mathcal{N}_n$ is, the more likely the edge state $\vert\psi_n(\theta)\rangle$ will break the adiabaticity, yielding excitation to the bulk state $\vert\psi_m(\theta)\rangle$ in the pumping process. One can check immediately that for the LZS model in Fig. \ref{fig-fig1} (c) and Eq. \eqref{eq-twolevel}, the non-adiabaticity $\mathcal{N}_n$ peaks at $t = T/4$ and $3T/4$. 

Thus the non-adiabaticity $\mathcal{N}_n$ encapsulates our central idea in a much clearer way. In regime A, $\mathcal{N}_n(\theta) \sim 1/\sqrt{L}$; and in regime B, 
$\mathcal{N}_n(\theta) \sim \text{constant}$. This is because in regime B, the level spacing goes as $1/\Delta \sim L$ and the overlap $O_{n,m} \sim 1/L$ since both wave functions are delocalized with amplitude $\sim 1/\sqrt{L}$. Thus the adiabatic condition is fulfilled by controlling the sweeping rate $d\theta/dt$, which is essential for the equivalent dynamics between Eq. \eqref{eq-Hamiltonian} and the LZS model. Moreover specifically at the NAPs, the values of non-adiabacity become very large. This picture is confirmed in Fig. \ref{fig-fig3}(d), in which the peaks of $\mathcal{N}_{26}(\theta)\simeq N_{26,27} $ are used to pinpoint the NAPs. 

\begin{figure}[hbtp!]
  \includegraphics[width=0.45\textwidth]{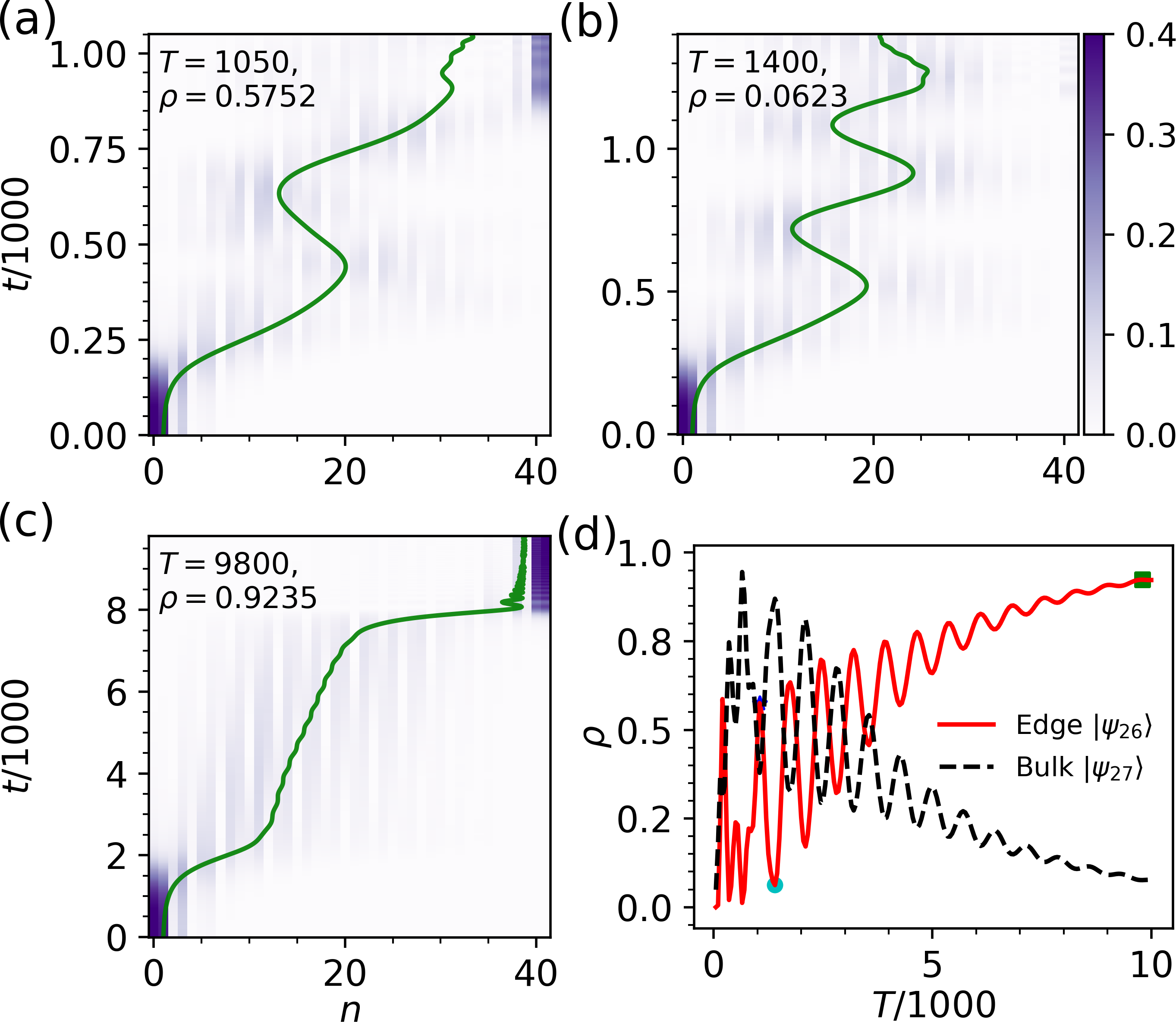}
  \caption {
  Distribution of wave function in spacetime and its mean trajectories (green curves) for three pump times: (a) $T=1050$, (b) $T=1400$ and (c) $T=9800$ with disorder strength $W = 0.08$ and $L=42$. (d) Occupation probability $\rho$ versus pump time $T$ for edge state $\vert \psi_{26}\rangle$ and bulk state $\vert \psi_{27}\rangle$. Note that the three pump times are chosen from three extrema of $\rho_{26}$, whose occupation probabilities are marked in (d). }
    \label{fig-fig4}
\end{figure}

\begin{figure}[hbtp!]
  \includegraphics[width=0.45\textwidth]{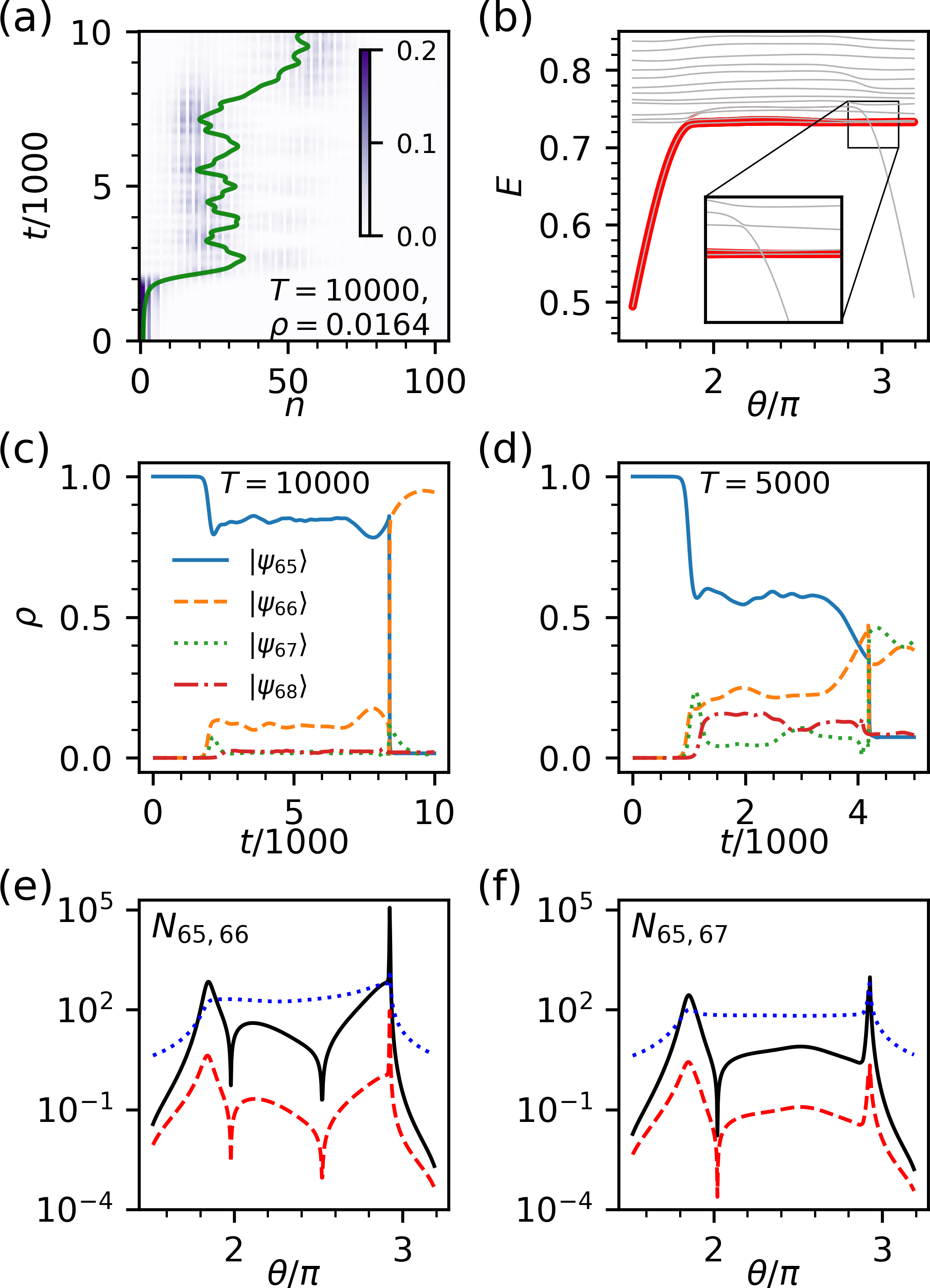}
  \caption{(a) Distribution of wave function in spacetime and its mean trajectories (solid lines) for pump time $T=10000$ with disorder strength $W = 0.08$ and $L = 105$. (b) Energy bands versus tuning phase $\theta$ where the occupation for edge state $\vert \psi_{65}\rangle$. (c-d) Occupation evolution $\rho(t)$ for edge state $\vert \psi_{65}\rangle$ and the three nearest bulk states above $\vert \psi_{66}\rangle$, $\vert \psi_{67}\rangle$ and $\vert \psi_{68}\rangle$ for (c) $T=10000$, and (d) $T=5000$ respectively. 
(e) Non-adiabaticity ${N}_{65,66}$ (solid line), overlapping $O_{65,66}$ (dashed line) and reciprocal energy gap $\vert \Delta_{65, 66}\vert^{-1}$ (dotted line), all the three versus tuning phase $\theta$. 
(f)  non-adiabaticity ${N}_{65,67}$ (solid line), overlapping $O_{65,67}$ (dashed line) and reciprocal energy gap $\vert \Delta_{65, 67}\vert^{-1}$ (dotted line), also versus $\theta$.  }
  \label{fig-fig5}
\end{figure}

{\it ESP with disorder}: The LZS interference and the non-adiabaticity at NAPs provide a unique way for us to understand the ESP with disorder, which is inevitably present in experiments \cite{Grabarits2024Floquet-Anderson}. In this case the wave function may become localized with localization length $\xi$  given by  \cite{Thouless1981Conductivity, Lin2024Fate} 
\begin{equation}
\xi^{-1} ={W^2 \over 24(4V^2 - E^2)},
\end{equation} 
where $V$ is the hopping strength, $E$ is the eigenvalue and $W$ is the disorder strength, all defined in Eq. \eqref{eq-Hamiltonian}. When $\xi \ll L$ with $L$ the chain length, the state with eigenvalues $E$ should be regarded as localized. In this case, the edge state, if existed, is impossible to be transferred from the left end to the right end, since the overlap between the wave functions, ${O}_{n,m}$, will become exponentially small. In the following, we focus on the case of $\xi \gg L$, where the bulk states are not fully localized. This condition will lead to intriguing physics not observed in the previous experiments. We first present the dynamics for a short chain $L =42$ in Fig. \ref{fig-fig4}, in which the behaviors of the results are similar to that of a clean chain in Fig. \ref{fig-fig2}. Surprisingly, with the size increased to $L = 105$, the dynamics is totally different. We find that even for a long pump time $T = 10000$, the left edge state cannot be transferred to the right end. The reason is presented in Fig. \ref{fig-fig5} (b), showing the anticrossing between the edge state and the bulk state at some $\theta$. We find that the edge state will gradually submerge into the bulk, and thus the ESP fails to happen for $T = 10^4$ and $T =5\times 10^3$ [see Fig. \ref{fig-fig5} (c-d)]. This failure is attributed to the non-adiabaticities for the edge state $\vert \psi_{65}\rangle$ with respect to its adjacent excited states $\vert\psi_{66}\rangle$ and $\vert\psi_{67}\rangle$: $N_{65, 66}$ and $N_{65, 67}$ in panels (e, f) respectively, which peak the most at the second NAP due to their anticrossing. The presence of anticrossing between the edge and bulk states implies a new mechanism for the non-adiabaticity in ESP. 

{\it Conclusion}: In this work we interpret the ESP in terms of LZS interference, which countertuitively breaks the adiabatic condition at two NAPs. This LZS interference at NAPs naturally explains the oscillation of ESP as a function of evolution time, and also contributes to its efficient pumping. In the disordered case, we reveal that the possible anticrossing between the edge and the bulk states even leads to failure of ESP. Our results show that the ESP available in the current experiments requires scrutinisation in much longer chains under more stringent conditions \cite{Rechtsman2013, ZhangY2018Resonant, Jurgensen2021, ChenZ2021}. We put forward an open question how one preserves the adiabatic condition during ESP, by engineering the driven protocol to achieve transport in a much shorter time \cite{Privitera2018Nonadiabatic, Citro2023, Longhi2019Topo}, which is essential for the manipulation of topological edge states in large, realistic systems. Our theoretical predictions can be simulated using ultracold atoms \cite{Schomerus2013Topo, Viebahn2019, Nakajima2021}, optical and acoustic waveguides \cite{XiaY2021, WangS2023}
and electric circuits \cite{Li2023Observation, Mei2018Robust}, in which the nonlinearity and non-Hermitian \cite{Jurgensen2021, LinQ2022, Wang2018Dynamical} can be naturally included. This unique platform for realizing of LZS model in our work, when using the various topological models with different symmetries  \cite{Qi2011, Hasan2010, Schnyder2008Classification}, will also greatly broaden the family of LZS systems for this intriguing dynamics, and unfold their associated nonadiabatic physics~\cite{Ivakhnenko2023Nonadiabatic}. 

{\it Acknowledgments}: Y. L. is supported by Young Scientist Fund [NSFC11804087], Natural National Science Foundation [NSFC12047501, 12074107 and 12075081]; and Science and Technology Department of Hubei Province [2022CFB553, 2022CFA012, 2024AFA038].  

%\end{acknowledgments}

%\nocite{*}

%\bibliography{ref9}

\begin{thebibliography}{82}%
\makeatletter
\providecommand \@ifxundefined [1]{%
 \@ifx{#1\undefined}
}%
\providecommand \@ifnum [1]{%
 \ifnum #1\expandafter \@firstoftwo
 \else \expandafter \@secondoftwo
 \fi
}%
\providecommand \@ifx [1]{%
 \ifx #1\expandafter \@firstoftwo
 \else \expandafter \@secondoftwo
 \fi
}%
\providecommand \natexlab [1]{#1}%
\providecommand \enquote  [1]{``#1''}%
\providecommand \bibnamefont  [1]{#1}%
\providecommand \bibfnamefont [1]{#1}%
\providecommand \citenamefont [1]{#1}%
\providecommand \href@noop [0]{\@secondoftwo}%
\providecommand \href [0]{\begingroup \@sanitize@url \@href}%
\providecommand \@href[1]{\@@startlink{#1}\@@href}%
\providecommand \@@href[1]{\endgroup#1\@@endlink}%
\providecommand \@sanitize@url [0]{\catcode `\\12\catcode `\$12\catcode
  `\&12\catcode `\#12\catcode `\^12\catcode `\_12\catcode `\%12\relax}%
\providecommand \@@startlink[1]{}%
\providecommand \@@endlink[0]{}%
\providecommand \url  [0]{\begingroup\@sanitize@url \@url }%
\providecommand \@url [1]{\endgroup\@href {#1}{\urlprefix }}%
\providecommand \urlprefix  [0]{URL }%
\providecommand \Eprint [0]{\href }%
\providecommand \doibase [0]{http://dx.doi.org/}%
\providecommand \selectlanguage [0]{\@gobble}%
\providecommand \bibinfo  [0]{\@secondoftwo}%
\providecommand \bibfield  [0]{\@secondoftwo}%
\providecommand \translation [1]{[#1]}%
\providecommand \BibitemOpen [0]{}%
\providecommand \bibitemStop [0]{}%
\providecommand \bibitemNoStop [0]{.\EOS\space}%
\providecommand \EOS [0]{\spacefactor3000\relax}%
\providecommand \BibitemShut  [1]{\csname bibitem#1\endcsname}%
\let\auto@bib@innerbib\@empty
%</preamble>
\bibitem [{\citenamefont {Qi}\ and\ \citenamefont {Zhang}(2011)}]{Qi2011}%
  \BibitemOpen
  \bibfield  {author} {\bibinfo {author} {\bibfnamefont {X.-L.}\ \bibnamefont
  {Qi}}\ and\ \bibinfo {author} {\bibfnamefont {S.-C.}\ \bibnamefont {Zhang}},\
  }\bibfield  {title} {\enquote {\bibinfo {title} {Topological insulators and
  superconductors},}\ }\href@noop {} {\bibfield  {journal} {\bibinfo  {journal}
  {Rev. Mod. Phys.}\ }\textbf {\bibinfo {volume} {83}},\ \bibinfo {pages}
  {1057} (\bibinfo {year} {2011})}\BibitemShut {NoStop}%
\bibitem [{\citenamefont {Hasan}\ and\ \citenamefont {Kane}(2010)}]{Hasan2010}%
  \BibitemOpen
  \bibfield  {author} {\bibinfo {author} {\bibfnamefont {M.~Z.}\ \bibnamefont
  {Hasan}}\ and\ \bibinfo {author} {\bibfnamefont {C.~L.}\ \bibnamefont
  {Kane}},\ }\bibfield  {title} {\enquote {\bibinfo {title} {Colloquium:
  topological insulators},}\ }\href@noop {} {\bibfield  {journal} {\bibinfo
  {journal} {Rev. Mod. Phys.}\ }\textbf {\bibinfo {volume} {82}},\ \bibinfo
  {pages} {3045} (\bibinfo {year} {2010})}\BibitemShut {NoStop}%
\bibitem [{\citenamefont {Schnyder}\ \emph {et~al.}(2008)\citenamefont
  {Schnyder}, \citenamefont {Ryu}, \citenamefont {Furusaki},\ and\
  \citenamefont {Ludwig}}]{Schnyder2008Classification}%
  \BibitemOpen
  \bibfield  {author} {\bibinfo {author} {\bibfnamefont {A.~P.}\ \bibnamefont
  {Schnyder}}, \bibinfo {author} {\bibfnamefont {S.}~\bibnamefont {Ryu}},
  \bibinfo {author} {\bibfnamefont {A.}~\bibnamefont {Furusaki}}, \ and\
  \bibinfo {author} {\bibfnamefont {A.~W.~W.}\ \bibnamefont {Ludwig}},\
  }\bibfield  {title} {\enquote {\bibinfo {title} {Classification of
  topological insulators and superconductors in three spatial dimensions},}\
  }\href {\doibase 10.1103/PhysRevB.78.195125} {\bibfield  {journal} {\bibinfo
  {journal} {Phys. Rev. B}\ }\textbf {\bibinfo {volume} {78}},\ \bibinfo
  {pages} {195125} (\bibinfo {year} {2008})}\BibitemShut {NoStop}%
\bibitem [{\citenamefont {Ozawa}\ \emph {et~al.}(2019)\citenamefont {Ozawa},
  \citenamefont {Price}, \citenamefont {Amo}, \citenamefont {Goldman},
  \citenamefont {Hafezi}, \citenamefont {Lu}, \citenamefont {Rechtsman},
  \citenamefont {Schuster}, \citenamefont {Simon}, \citenamefont {Zilberberg}
  \emph {et~al.}}]{Ozawa2019a}%
  \BibitemOpen
  \bibfield  {author} {\bibinfo {author} {\bibfnamefont {T.}~\bibnamefont
  {Ozawa}}, \bibinfo {author} {\bibfnamefont {H.~M.}\ \bibnamefont {Price}},
  \bibinfo {author} {\bibfnamefont {A.}~\bibnamefont {Amo}}, \bibinfo {author}
  {\bibfnamefont {N.}~\bibnamefont {Goldman}}, \bibinfo {author} {\bibfnamefont
  {M.}~\bibnamefont {Hafezi}}, \bibinfo {author} {\bibfnamefont
  {L.}~\bibnamefont {Lu}}, \bibinfo {author} {\bibfnamefont {M.~C.}\
  \bibnamefont {Rechtsman}}, \bibinfo {author} {\bibfnamefont {D.}~\bibnamefont
  {Schuster}}, \bibinfo {author} {\bibfnamefont {J.}~\bibnamefont {Simon}},
  \bibinfo {author} {\bibfnamefont {O.}~\bibnamefont {Zilberberg}},  \emph
  {et~al.},\ }\bibfield  {title} {\enquote {\bibinfo {title} {Topological
  photonics},}\ }\href@noop {} {\bibfield  {journal} {\bibinfo  {journal} {Rev.
  Mod. Phys.}\ }\textbf {\bibinfo {volume} {91}},\ \bibinfo {pages} {015006}
  (\bibinfo {year} {2019})}\BibitemShut {NoStop}%
\bibitem [{\citenamefont {Georgescu}\ \emph {et~al.}(2014)\citenamefont
  {Georgescu}, \citenamefont {Ashhab},\ and\ \citenamefont
  {Nori}}]{Georgescu2014Quantum}%
  \BibitemOpen
  \bibfield  {author} {\bibinfo {author} {\bibfnamefont {I.~M.}\ \bibnamefont
  {Georgescu}}, \bibinfo {author} {\bibfnamefont {S.}~\bibnamefont {Ashhab}}, \
  and\ \bibinfo {author} {\bibfnamefont {F.}~\bibnamefont {Nori}},\ }\bibfield
  {title} {\enquote {\bibinfo {title} {Quantum simulation},}\ }\href {\doibase
  10.1103/RevModPhys.86.153} {\bibfield  {journal} {\bibinfo  {journal} {Rev.
  Mod. Phys.}\ }\textbf {\bibinfo {volume} {86}},\ \bibinfo {pages} {153--185}
  (\bibinfo {year} {2014})}\BibitemShut {NoStop}%
\bibitem [{\citenamefont {Thouless}(1983)}]{Thouless1983Quantization}%
  \BibitemOpen
  \bibfield  {author} {\bibinfo {author} {\bibfnamefont {D.~J.}\ \bibnamefont
  {Thouless}},\ }\bibfield  {title} {\enquote {\bibinfo {title} {Quantization
  of particle transport},}\ }\href {\doibase 10.1103/PhysRevB.27.6083}
  {\bibfield  {journal} {\bibinfo  {journal} {Phys. Rev. B}\ }\textbf {\bibinfo
  {volume} {27}},\ \bibinfo {pages} {6083--6087} (\bibinfo {year}
  {1983})}\BibitemShut {NoStop}%
\bibitem [{\citenamefont {Citro}\ and\ \citenamefont
  {Aidelsburger}(2023)}]{Citro2023}%
  \BibitemOpen
  \bibfield  {author} {\bibinfo {author} {\bibfnamefont {R.}~\bibnamefont
  {Citro}}\ and\ \bibinfo {author} {\bibfnamefont {M.}~\bibnamefont
  {Aidelsburger}},\ }\bibfield  {title} {\enquote {\bibinfo {title} {Thouless
  pumping and topology},}\ }\href {\doibase 10.1038/s42254-022-00545-0}
  {\bibfield  {journal} {\bibinfo  {journal} {Nat. Rev. Phys.}\ }\textbf
  {\bibinfo {volume} {5}},\ \bibinfo {pages} {87--101} (\bibinfo {year}
  {2023})}\BibitemShut {NoStop}%
\bibitem [{\citenamefont {Aleiner}\ and\ \citenamefont
  {Andreev}(1998)}]{Aleiner1998Adiabatic}%
  \BibitemOpen
  \bibfield  {author} {\bibinfo {author} {\bibfnamefont {I.~L.}\ \bibnamefont
  {Aleiner}}\ and\ \bibinfo {author} {\bibfnamefont {A.~V.}\ \bibnamefont
  {Andreev}},\ }\bibfield  {title} {\enquote {\bibinfo {title} {Adiabatic
  charge pumping in almost open dots},}\ }\href {\doibase
  10.1103/PhysRevLett.81.1286} {\bibfield  {journal} {\bibinfo  {journal}
  {Phys. Rev. Lett.}\ }\textbf {\bibinfo {volume} {81}},\ \bibinfo {pages}
  {1286--1289} (\bibinfo {year} {1998})}\BibitemShut {NoStop}%
\bibitem [{\citenamefont {Nakajima}\ \emph {et~al.}(2016)\citenamefont
  {Nakajima}, \citenamefont {Tomita}, \citenamefont {Taie}, \citenamefont
  {Ichinose}, \citenamefont {Ozawa}, \citenamefont {Wang}, \citenamefont
  {Troyer},\ and\ \citenamefont {Takahashi}}]{Nakajima2016Topo}%
  \BibitemOpen
  \bibfield  {author} {\bibinfo {author} {\bibfnamefont {S.}~\bibnamefont
  {Nakajima}}, \bibinfo {author} {\bibfnamefont {T.}~\bibnamefont {Tomita}},
  \bibinfo {author} {\bibfnamefont {S.}~\bibnamefont {Taie}}, \bibinfo {author}
  {\bibfnamefont {T.}~\bibnamefont {Ichinose}}, \bibinfo {author}
  {\bibfnamefont {H.}~\bibnamefont {Ozawa}}, \bibinfo {author} {\bibfnamefont
  {L.}~\bibnamefont {Wang}}, \bibinfo {author} {\bibfnamefont {M.}~\bibnamefont
  {Troyer}}, \ and\ \bibinfo {author} {\bibfnamefont {Y.}~\bibnamefont
  {Takahashi}},\ }\bibfield  {title} {\enquote {\bibinfo {title} {Topological
  thouless pumping of ultracold fermions},}\ }\href {\doibase
  10.1038/nphys3622} {\bibfield  {journal} {\bibinfo  {journal} {Nat. Phys.}\
  }\textbf {\bibinfo {volume} {12}},\ \bibinfo {pages} {296--300} (\bibinfo
  {year} {2016})}\BibitemShut {NoStop}%
\bibitem [{\citenamefont {Sun}\ \emph {et~al.}(2022)\citenamefont {Sun},
  \citenamefont {Zhang}, \citenamefont {Yu}, \citenamefont {Tian},
  \citenamefont {Chen},\ and\ \citenamefont {Sun}}]{SunY2022Non-Abelian}%
  \BibitemOpen
  \bibfield  {author} {\bibinfo {author} {\bibfnamefont {Y.-K.}\ \bibnamefont
  {Sun}}, \bibinfo {author} {\bibfnamefont {X.-L.}\ \bibnamefont {Zhang}},
  \bibinfo {author} {\bibfnamefont {F.}~\bibnamefont {Yu}}, \bibinfo {author}
  {\bibfnamefont {Z.-N.}\ \bibnamefont {Tian}}, \bibinfo {author}
  {\bibfnamefont {Q.-D.}\ \bibnamefont {Chen}}, \ and\ \bibinfo {author}
  {\bibfnamefont {H.-B.}\ \bibnamefont {Sun}},\ }\bibfield  {title} {\enquote
  {\bibinfo {title} {Non-{Abelian Thouless} pumping in photonic waveguides},}\
  }\href {\doibase 10.1038/s41567-022-01669-x} {\bibfield  {journal} {\bibinfo
  {journal} {Nat. Phys.}\ }\textbf {\bibinfo {volume} {18}},\ \bibinfo {pages}
  {1080--1085} (\bibinfo {year} {2022})}\BibitemShut {NoStop}%
\bibitem [{\citenamefont {J\"urgensen}\ \emph {et~al.}(2021)\citenamefont
  {J\"urgensen}, \citenamefont {Mukherjee},\ and\ \citenamefont
  {Rechtsman}}]{Jurgensen2021}%
  \BibitemOpen
  \bibfield  {author} {\bibinfo {author} {\bibfnamefont {M.}~\bibnamefont
  {J\"urgensen}}, \bibinfo {author} {\bibfnamefont {S.}~\bibnamefont
  {Mukherjee}}, \ and\ \bibinfo {author} {\bibfnamefont {M.~C.}\ \bibnamefont
  {Rechtsman}},\ }\bibfield  {title} {\enquote {\bibinfo {title} {Quantized
  nonlinear {Thouless} pumping},}\ }\href {\doibase 10.1038/s41586-021-03688-9}
  {\bibfield  {journal} {\bibinfo  {journal} {Nature}\ }\textbf {\bibinfo
  {volume} {596}},\ \bibinfo {pages} {63--67} (\bibinfo {year}
  {2021})}\BibitemShut {NoStop}%
\bibitem [{\citenamefont {J\"urgensen}\ \emph {et~al.}(2023)\citenamefont
  {J\"urgensen}, \citenamefont {Mukherjee}, \citenamefont {J\"org},\ and\
  \citenamefont {Rechtsman}}]{Jurgensen2023}%
  \BibitemOpen
  \bibfield  {author} {\bibinfo {author} {\bibfnamefont {M.}~\bibnamefont
  {J\"urgensen}}, \bibinfo {author} {\bibfnamefont {S.}~\bibnamefont
  {Mukherjee}}, \bibinfo {author} {\bibfnamefont {C.}~\bibnamefont {J\"org}}, \
  and\ \bibinfo {author} {\bibfnamefont {M.~C.}\ \bibnamefont {Rechtsman}},\
  }\bibfield  {title} {\enquote {\bibinfo {title} {Quantized fractional
  {Thouless} pumping of solitons},}\ }\href {\doibase
  10.1038/s41567-022-01871-x} {\bibfield  {journal} {\bibinfo  {journal} {Nat.
  Phys.}\ }\textbf {\bibinfo {volume} {19}},\ \bibinfo {pages} {420--426}
  (\bibinfo {year} {2023})}\BibitemShut {NoStop}%
\bibitem [{\citenamefont {Cerjan}\ \emph {et~al.}(2020)\citenamefont {Cerjan},
  \citenamefont {Wang}, \citenamefont {Huang}, \citenamefont {Chen},\ and\
  \citenamefont {Rechtsman}}]{Cerjan2020}%
  \BibitemOpen
  \bibfield  {author} {\bibinfo {author} {\bibfnamefont {A.}~\bibnamefont
  {Cerjan}}, \bibinfo {author} {\bibfnamefont {M.}~\bibnamefont {Wang}},
  \bibinfo {author} {\bibfnamefont {S.}~\bibnamefont {Huang}}, \bibinfo
  {author} {\bibfnamefont {K.~P.}\ \bibnamefont {Chen}}, \ and\ \bibinfo
  {author} {\bibfnamefont {M.~C.}\ \bibnamefont {Rechtsman}},\ }\bibfield
  {title} {\enquote {\bibinfo {title} {Thouless pumping in disordered photonic
  systems},}\ }\href {\doibase 10.1038/s41377-020-00408-2} {\bibfield
  {journal} {\bibinfo  {journal} {Light Sci. Appl.}\ }\textbf {\bibinfo
  {volume} {9}},\ \bibinfo {pages} {178} (\bibinfo {year} {2020})}\BibitemShut
  {NoStop}%
\bibitem [{\citenamefont {Lahini}\ \emph {et~al.}(2009)\citenamefont {Lahini},
  \citenamefont {Pugatch}, \citenamefont {Pozzi}, \citenamefont {Sorel},
  \citenamefont {Morandotti}, \citenamefont {Davidson},\ and\ \citenamefont
  {Silberberg}}]{Lahini2009}%
  \BibitemOpen
  \bibfield  {author} {\bibinfo {author} {\bibfnamefont {Y.}~\bibnamefont
  {Lahini}}, \bibinfo {author} {\bibfnamefont {R.}~\bibnamefont {Pugatch}},
  \bibinfo {author} {\bibfnamefont {F.}~\bibnamefont {Pozzi}}, \bibinfo
  {author} {\bibfnamefont {M.}~\bibnamefont {Sorel}}, \bibinfo {author}
  {\bibfnamefont {R.}~\bibnamefont {Morandotti}}, \bibinfo {author}
  {\bibfnamefont {N.}~\bibnamefont {Davidson}}, \ and\ \bibinfo {author}
  {\bibfnamefont {Y.}~\bibnamefont {Silberberg}},\ }\bibfield  {title}
  {\enquote {\bibinfo {title} {Observation of a localization transition in
  quasiperiodic photonic lattices},}\ }\href {\doibase
  10.1103/PhysRevLett.103.013901} {\bibfield  {journal} {\bibinfo  {journal}
  {Phys. Rev. Lett.}\ }\textbf {\bibinfo {volume} {103}},\ \bibinfo {pages}
  {013901} (\bibinfo {year} {2009})}\BibitemShut {NoStop}%
\bibitem [{\citenamefont {Kraus}\ \emph {et~al.}(2012)\citenamefont {Kraus},
  \citenamefont {Lahini}, \citenamefont {Ringel}, \citenamefont {Verbin},\ and\
  \citenamefont {Zilberberg}}]{Kraus2012}%
  \BibitemOpen
  \bibfield  {author} {\bibinfo {author} {\bibfnamefont {Y.~E.}\ \bibnamefont
  {Kraus}}, \bibinfo {author} {\bibfnamefont {Y.}~\bibnamefont {Lahini}},
  \bibinfo {author} {\bibfnamefont {Z.}~\bibnamefont {Ringel}}, \bibinfo
  {author} {\bibfnamefont {M.}~\bibnamefont {Verbin}}, \ and\ \bibinfo {author}
  {\bibfnamefont {O.}~\bibnamefont {Zilberberg}},\ }\bibfield  {title}
  {\enquote {\bibinfo {title} {Topological states and adiabatic pumping in
  quasicrystals},}\ }\href {\doibase 10.1103/PhysRevLett.109.106402} {\bibfield
   {journal} {\bibinfo  {journal} {Phys. Rev. Lett.}\ }\textbf {\bibinfo
  {volume} {109}},\ \bibinfo {pages} {106402} (\bibinfo {year}
  {2012})}\BibitemShut {NoStop}%
\bibitem [{\citenamefont {Rechtsman}\ \emph {et~al.}(2013)\citenamefont
  {Rechtsman}, \citenamefont {Zeuner}, \citenamefont {Plotnik}, \citenamefont
  {Lumer}, \citenamefont {Podolsky}, \citenamefont {Dreisow}, \citenamefont
  {Nolte}, \citenamefont {Segev},\ and\ \citenamefont
  {Szameit}}]{Rechtsman2013}%
  \BibitemOpen
  \bibfield  {author} {\bibinfo {author} {\bibfnamefont {M.~C.}\ \bibnamefont
  {Rechtsman}}, \bibinfo {author} {\bibfnamefont {J.~M.}\ \bibnamefont
  {Zeuner}}, \bibinfo {author} {\bibfnamefont {Y.}~\bibnamefont {Plotnik}},
  \bibinfo {author} {\bibfnamefont {Y.}~\bibnamefont {Lumer}}, \bibinfo
  {author} {\bibfnamefont {D.}~\bibnamefont {Podolsky}}, \bibinfo {author}
  {\bibfnamefont {F.}~\bibnamefont {Dreisow}}, \bibinfo {author} {\bibfnamefont
  {S.}~\bibnamefont {Nolte}}, \bibinfo {author} {\bibfnamefont
  {M.}~\bibnamefont {Segev}}, \ and\ \bibinfo {author} {\bibfnamefont
  {A.}~\bibnamefont {Szameit}},\ }\bibfield  {title} {\enquote {\bibinfo
  {title} {Photonic {Floquet} topological insulators},}\ }\href {\doibase
  10.1038/nature12066} {\bibfield  {journal} {\bibinfo  {journal} {Nature}\
  }\textbf {\bibinfo {volume} {496}},\ \bibinfo {pages} {196--200} (\bibinfo
  {year} {2013})}\BibitemShut {NoStop}%
\bibitem [{\citenamefont {Tambasco}\ \emph {et~al.}(2018)\citenamefont
  {Tambasco}, \citenamefont {Corrielli}, \citenamefont {Chapman}, \citenamefont
  {Crespi}, \citenamefont {Zilberberg}, \citenamefont {Osellame},\ and\
  \citenamefont {Peruzzo}}]{Tambasco2018Quantum}%
  \BibitemOpen
  \bibfield  {author} {\bibinfo {author} {\bibfnamefont {J.-L.}\ \bibnamefont
  {Tambasco}}, \bibinfo {author} {\bibfnamefont {G.}~\bibnamefont {Corrielli}},
  \bibinfo {author} {\bibfnamefont {R.~J.}\ \bibnamefont {Chapman}}, \bibinfo
  {author} {\bibfnamefont {A.}~\bibnamefont {Crespi}}, \bibinfo {author}
  {\bibfnamefont {O.}~\bibnamefont {Zilberberg}}, \bibinfo {author}
  {\bibfnamefont {R.}~\bibnamefont {Osellame}}, \ and\ \bibinfo {author}
  {\bibfnamefont {A.}~\bibnamefont {Peruzzo}},\ }\bibfield  {title} {\enquote
  {\bibinfo {title} {Quantum interference of topological states of light},}\
  }\href {\doibase 10.1126/sciadv.aat3187} {\bibfield  {journal} {\bibinfo
  {journal} {Science Advances}\ }\textbf {\bibinfo {volume} {4}},\ \bibinfo
  {pages} {eaat3187} (\bibinfo {year} {2018})}\BibitemShut {NoStop}%
\bibitem [{\citenamefont {Wang}\ \emph {et~al.}(2019)\citenamefont {Wang},
  \citenamefont {Pang}, \citenamefont {Lu}, \citenamefont {Gao}, \citenamefont
  {Chang}, \citenamefont {Qiao}, \citenamefont {Jiao}, \citenamefont {Tang},\
  and\ \citenamefont {Jin}}]{WangY2019}%
  \BibitemOpen
  \bibfield  {author} {\bibinfo {author} {\bibfnamefont {Yao}\ \bibnamefont
  {Wang}}, \bibinfo {author} {\bibfnamefont {Xiao-Ling}\ \bibnamefont {Pang}},
  \bibinfo {author} {\bibfnamefont {Yong-Heng}\ \bibnamefont {Lu}}, \bibinfo
  {author} {\bibfnamefont {Jun}\ \bibnamefont {Gao}}, \bibinfo {author}
  {\bibfnamefont {Yi-Jun}\ \bibnamefont {Chang}}, \bibinfo {author}
  {\bibfnamefont {Lu-Feng}\ \bibnamefont {Qiao}}, \bibinfo {author}
  {\bibfnamefont {Zhi-Qiang}\ \bibnamefont {Jiao}}, \bibinfo {author}
  {\bibfnamefont {Hao}\ \bibnamefont {Tang}}, \ and\ \bibinfo {author}
  {\bibfnamefont {Xian-Min}\ \bibnamefont {Jin}},\ }\bibfield  {title}
  {\enquote {\bibinfo {title} {Topological protection of two-photon quantum
  correlation on a photonic chip},}\ }\href {\doibase 10.1364/optica.6.000955}
  {\bibfield  {journal} {\bibinfo  {journal} {Optica}\ }\textbf {\bibinfo
  {volume} {6}} (\bibinfo {year} {2019}),\ 10.1364/optica.6.000955}\BibitemShut
  {NoStop}%
\bibitem [{\citenamefont {Yuan}\ \emph {et~al.}(2018)\citenamefont {Yuan},
  \citenamefont {Lin}, \citenamefont {Xiao},\ and\ \citenamefont
  {Fan}}]{YuanL2018}%
  \BibitemOpen
  \bibfield  {author} {\bibinfo {author} {\bibfnamefont {Luqi}\ \bibnamefont
  {Yuan}}, \bibinfo {author} {\bibfnamefont {Qian}\ \bibnamefont {Lin}},
  \bibinfo {author} {\bibfnamefont {Meng}\ \bibnamefont {Xiao}}, \ and\
  \bibinfo {author} {\bibfnamefont {Shanhui}\ \bibnamefont {Fan}},\ }\bibfield
  {title} {\enquote {\bibinfo {title} {Synthetic dimension in photonics},}\
  }\href {\doibase 10.1364/optica.5.001396} {\bibfield  {journal} {\bibinfo
  {journal} {Optica}\ }\textbf {\bibinfo {volume} {5}} (\bibinfo {year}
  {2018}),\ 10.1364/optica.5.001396}\BibitemShut {NoStop}%
\bibitem [{\citenamefont {Zilberberg}\ \emph {et~al.}(2018)\citenamefont
  {Zilberberg}, \citenamefont {Huang}, \citenamefont {Guglielmon},
  \citenamefont {Wang}, \citenamefont {Chen}, \citenamefont {Kraus},\ and\
  \citenamefont {Rechtsman}}]{Zilberberg2018}%
  \BibitemOpen
  \bibfield  {author} {\bibinfo {author} {\bibfnamefont {O.}~\bibnamefont
  {Zilberberg}}, \bibinfo {author} {\bibfnamefont {S.}~\bibnamefont {Huang}},
  \bibinfo {author} {\bibfnamefont {J.}~\bibnamefont {Guglielmon}}, \bibinfo
  {author} {\bibfnamefont {M.}~\bibnamefont {Wang}}, \bibinfo {author}
  {\bibfnamefont {K.~P.}\ \bibnamefont {Chen}}, \bibinfo {author}
  {\bibfnamefont {Y.~E.}\ \bibnamefont {Kraus}}, \ and\ \bibinfo {author}
  {\bibfnamefont {M.~C.}\ \bibnamefont {Rechtsman}},\ }\bibfield  {title}
  {\enquote {\bibinfo {title} {Photonic topological boundary pumping as a probe
  of {4D} quantum {Hall} physics},}\ }\href {\doibase 10.1038/nature25011}
  {\bibfield  {journal} {\bibinfo  {journal} {Nature}\ }\textbf {\bibinfo
  {volume} {553}},\ \bibinfo {pages} {59--62} (\bibinfo {year}
  {2018})}\BibitemShut {NoStop}%
\bibitem [{\citenamefont {Yang}\ \emph {et~al.}(2020)\citenamefont {Yang},
  \citenamefont {Lustig}, \citenamefont {Harari}, \citenamefont {Plotnik},
  \citenamefont {Lumer}, \citenamefont {Bandres},\ and\ \citenamefont
  {Segev}}]{YangZ2020}%
  \BibitemOpen
  \bibfield  {author} {\bibinfo {author} {\bibfnamefont {Z.}~\bibnamefont
  {Yang}}, \bibinfo {author} {\bibfnamefont {E.}~\bibnamefont {Lustig}},
  \bibinfo {author} {\bibfnamefont {G.}~\bibnamefont {Harari}}, \bibinfo
  {author} {\bibfnamefont {Y.}~\bibnamefont {Plotnik}}, \bibinfo {author}
  {\bibfnamefont {Y.}~\bibnamefont {Lumer}}, \bibinfo {author} {\bibfnamefont
  {M.~A.}\ \bibnamefont {Bandres}}, \ and\ \bibinfo {author} {\bibfnamefont
  {M.}~\bibnamefont {Segev}},\ }\bibfield  {title} {\enquote {\bibinfo {title}
  {Mode-locked topological insulator laser utilizing synthetic dimensions},}\
  }\href {\doibase 10.1103/PhysRevX.10.011059} {\bibfield  {journal} {\bibinfo
  {journal} {Phys. Rev. X}\ }\textbf {\bibinfo {volume} {10}},\ \bibinfo
  {pages} {011059} (\bibinfo {year} {2020})}\BibitemShut {NoStop}%
\bibitem [{\citenamefont {Nakajima}\ \emph {et~al.}(2021)\citenamefont
  {Nakajima}, \citenamefont {Takei}, \citenamefont {Kuno}, \citenamefont
  {Marra},\ and\ \citenamefont {Takahashi}}]{Nakajima2021}%
  \BibitemOpen
  \bibfield  {author} {\bibinfo {author} {\bibfnamefont {S.}~\bibnamefont
  {Nakajima}}, \bibinfo {author} {\bibfnamefont {K.}~\bibnamefont {Takei},
  \bibfnamefont {N.and~Sakuma}}, \bibinfo {author} {\bibfnamefont
  {Y.}~\bibnamefont {Kuno}}, \bibinfo {author} {\bibfnamefont {P.}~\bibnamefont
  {Marra}}, \ and\ \bibinfo {author} {\bibfnamefont {Y.}~\bibnamefont
  {Takahashi}},\ }\bibfield  {title} {\enquote {\bibinfo {title} {Competition
  and interplay between topology and quasi-periodic disorder in {Thouless}
  pumping of ultracold atoms},}\ }\href {\doibase 10.1038/s41567-021-01229-9}
  {\bibfield  {journal} {\bibinfo  {journal} {Nat. Phys.}\ }\textbf {\bibinfo
  {volume} {17}},\ \bibinfo {pages} {844--849} (\bibinfo {year}
  {2021})}\BibitemShut {NoStop}%
\bibitem [{\citenamefont {Kolodrubetz}\ \emph {et~al.}(2018)\citenamefont
  {Kolodrubetz}, \citenamefont {Nathan}, \citenamefont {Gazit}, \citenamefont
  {Morimoto},\ and\ \citenamefont {Moore}}]{Kolodrubet2018Floquet-Thouless}%
  \BibitemOpen
  \bibfield  {author} {\bibinfo {author} {\bibfnamefont {M.~H.}\ \bibnamefont
  {Kolodrubetz}}, \bibinfo {author} {\bibfnamefont {F.}~\bibnamefont {Nathan}},
  \bibinfo {author} {\bibfnamefont {S.}~\bibnamefont {Gazit}}, \bibinfo
  {author} {\bibfnamefont {T.}~\bibnamefont {Morimoto}}, \ and\ \bibinfo
  {author} {\bibfnamefont {J.~E.}\ \bibnamefont {Moore}},\ }\bibfield  {title}
  {\enquote {\bibinfo {title} {Topological {Floquet-Thouless} energy pump},}\
  }\href {\doibase 10.1103/PhysRevLett.120.150601} {\bibfield  {journal}
  {\bibinfo  {journal} {Phys. Rev. Lett.}\ }\textbf {\bibinfo {volume} {120}},\
  \bibinfo {pages} {150601} (\bibinfo {year} {2018})}\BibitemShut {NoStop}%
\bibitem [{\citenamefont {Tangpanitanon}\ \emph {et~al.}(2016)\citenamefont
  {Tangpanitanon}, \citenamefont {Bastidas}, \citenamefont {Al-Assam},
  \citenamefont {Roushan}, \citenamefont {Jaksch},\ and\ \citenamefont
  {Angelakis}}]{Tangpanitanon2016}%
  \BibitemOpen
  \bibfield  {author} {\bibinfo {author} {\bibfnamefont {J.}~\bibnamefont
  {Tangpanitanon}}, \bibinfo {author} {\bibfnamefont {V.~M.}\ \bibnamefont
  {Bastidas}}, \bibinfo {author} {\bibfnamefont {S.}~\bibnamefont {Al-Assam}},
  \bibinfo {author} {\bibfnamefont {P.}~\bibnamefont {Roushan}}, \bibinfo
  {author} {\bibfnamefont {D.}~\bibnamefont {Jaksch}}, \ and\ \bibinfo {author}
  {\bibfnamefont {D.~G.}\ \bibnamefont {Angelakis}},\ }\bibfield  {title}
  {\enquote {\bibinfo {title} {Topological pumping of photons in nonlinear
  resonator arrays},}\ }\href {\doibase 10.1103/PhysRevLett.117.213603}
  {\bibfield  {journal} {\bibinfo  {journal} {Phys. Rev. Lett.}\ }\textbf
  {\bibinfo {volume} {117}},\ \bibinfo {pages} {213603} (\bibinfo {year}
  {2016})}\BibitemShut {NoStop}%
\bibitem [{\citenamefont {Mostaan}\ \emph {et~al.}(2022)\citenamefont
  {Mostaan}, \citenamefont {Grusdt},\ and\ \citenamefont
  {Goldman}}]{Mostaan2022}%
  \BibitemOpen
  \bibfield  {author} {\bibinfo {author} {\bibfnamefont {N.}~\bibnamefont
  {Mostaan}}, \bibinfo {author} {\bibfnamefont {F.}~\bibnamefont {Grusdt}}, \
  and\ \bibinfo {author} {\bibfnamefont {N.}~\bibnamefont {Goldman}},\
  }\bibfield  {title} {\enquote {\bibinfo {title} {Quantized topological
  pumping of solitons in nonlinear photonics and ultracold atomic mixtures},}\
  }\href {\doibase 10.1038/s41467-022-33478-4} {\bibfield  {journal} {\bibinfo
  {journal} {Nat. Commun.}\ }\textbf {\bibinfo {volume} {13}},\ \bibinfo
  {pages} {5997} (\bibinfo {year} {2022})}\BibitemShut {NoStop}%
\bibitem [{\citenamefont {Qian}\ \emph {et~al.}(2011)\citenamefont {Qian},
  \citenamefont {Gong},\ and\ \citenamefont {Zhang}}]{Qian2011Quantum}%
  \BibitemOpen
  \bibfield  {author} {\bibinfo {author} {\bibfnamefont {Yinyin}\ \bibnamefont
  {Qian}}, \bibinfo {author} {\bibfnamefont {Ming}\ \bibnamefont {Gong}}, \
  and\ \bibinfo {author} {\bibfnamefont {Chuanwei}\ \bibnamefont {Zhang}},\
  }\bibfield  {title} {\enquote {\bibinfo {title} {Quantum transport of bosonic
  cold atoms in double-well optical lattices},}\ }\href {\doibase
  10.1103/PhysRevA.84.013608} {\bibfield  {journal} {\bibinfo  {journal} {Phys.
  Rev. A}\ }\textbf {\bibinfo {volume} {84}},\ \bibinfo {pages} {013608}
  (\bibinfo {year} {2011})}\BibitemShut {NoStop}%
\bibitem [{\citenamefont {Xia}\ \emph {et~al.}(2021)\citenamefont {Xia},
  \citenamefont {Riva}, \citenamefont {Rosa}, \citenamefont {Cazzulani},
  \citenamefont {Erturk}, \citenamefont {Braghin},\ and\ \citenamefont
  {Ruzzene}}]{XiaY2021}%
  \BibitemOpen
  \bibfield  {author} {\bibinfo {author} {\bibfnamefont {Y.}~\bibnamefont
  {Xia}}, \bibinfo {author} {\bibfnamefont {E.}~\bibnamefont {Riva}}, \bibinfo
  {author} {\bibfnamefont {M.~I.~N.}\ \bibnamefont {Rosa}}, \bibinfo {author}
  {\bibfnamefont {G.}~\bibnamefont {Cazzulani}}, \bibinfo {author}
  {\bibfnamefont {A.}~\bibnamefont {Erturk}}, \bibinfo {author} {\bibfnamefont
  {F.}~\bibnamefont {Braghin}}, \ and\ \bibinfo {author} {\bibfnamefont
  {M.}~\bibnamefont {Ruzzene}},\ }\bibfield  {title} {\enquote {\bibinfo
  {title} {Experimental observation of temporal pumping in electromechanical
  waveguides},}\ }\href {\doibase 10.1103/PhysRevLett.126.095501} {\bibfield
  {journal} {\bibinfo  {journal} {Phys. Rev. Lett.}\ }\textbf {\bibinfo
  {volume} {126}},\ \bibinfo {pages} {095501} (\bibinfo {year}
  {2021})}\BibitemShut {NoStop}%
\bibitem [{\citenamefont {Wang}\ \emph
  {et~al.}(2023{\natexlab{a}})\citenamefont {Wang}, \citenamefont {Hu},
  \citenamefont {Wu}, \citenamefont {Chen}, \citenamefont {Prodan},
  \citenamefont {Zhu},\ and\ \citenamefont {Huang}}]{WangS2023}%
  \BibitemOpen
  \bibfield  {author} {\bibinfo {author} {\bibfnamefont {S.}~\bibnamefont
  {Wang}}, \bibinfo {author} {\bibfnamefont {Z.}~\bibnamefont {Hu}}, \bibinfo
  {author} {\bibfnamefont {Q.}~\bibnamefont {Wu}}, \bibinfo {author}
  {\bibfnamefont {H.}~\bibnamefont {Chen}}, \bibinfo {author} {\bibfnamefont
  {E.}~\bibnamefont {Prodan}}, \bibinfo {author} {\bibfnamefont
  {R.}~\bibnamefont {Zhu}}, \ and\ \bibinfo {author} {\bibfnamefont
  {G.}~\bibnamefont {Huang}},\ }\bibfield  {title} {\enquote {\bibinfo {title}
  {Smart patterning for topological pumping of elastic surface waves},}\
  }\href@noop {} {\bibfield  {journal} {\bibinfo  {journal} {Science Advances}\
  }\textbf {\bibinfo {volume} {9}} (\bibinfo {year}
  {2023}{\natexlab{a}})}\BibitemShut {NoStop}%
\bibitem [{\citenamefont {Rosa}\ \emph {et~al.}(2019)\citenamefont {Rosa},
  \citenamefont {Pal}, \citenamefont {Arruda},\ and\ \citenamefont
  {Ruzzene}}]{Rosa2019}%
  \BibitemOpen
  \bibfield  {author} {\bibinfo {author} {\bibfnamefont {M.~I.~N.}\
  \bibnamefont {Rosa}}, \bibinfo {author} {\bibfnamefont {R.~K.}\ \bibnamefont
  {Pal}}, \bibinfo {author} {\bibfnamefont {J.~R.~F.}\ \bibnamefont {Arruda}},
  \ and\ \bibinfo {author} {\bibfnamefont {M.}~\bibnamefont {Ruzzene}},\
  }\bibfield  {title} {\enquote {\bibinfo {title} {Edge states and topological
  pumping in spatially modulated elastic lattices},}\ }\href {\doibase
  10.1103/PhysRevLett.123.034301} {\bibfield  {journal} {\bibinfo  {journal}
  {Phys. Rev. Lett.}\ }\textbf {\bibinfo {volume} {123}},\ \bibinfo {pages}
  {034301} (\bibinfo {year} {2019})}\BibitemShut {NoStop}%
\bibitem [{\citenamefont {Frank}\ \emph {et~al.}(2022)\citenamefont {Frank},
  \citenamefont {Leykam}, \citenamefont {Smirnova}, \citenamefont {Angelakis},\
  and\ \citenamefont {Ling}}]{Frank2022Boosting}%
  \BibitemOpen
  \bibfield  {author} {\bibinfo {author} {\bibfnamefont {A.}~\bibnamefont
  {Frank}}, \bibinfo {author} {\bibfnamefont {D.}~\bibnamefont {Leykam}},
  \bibinfo {author} {\bibfnamefont {D.~A.}\ \bibnamefont {Smirnova}}, \bibinfo
  {author} {\bibfnamefont {D.~G.}\ \bibnamefont {Angelakis}}, \ and\ \bibinfo
  {author} {\bibfnamefont {A.}~\bibnamefont {Ling}},\ }\bibfield  {title}
  {\enquote {\bibinfo {title} {Boosting topological zero modes using elastomer
  waveguide arrays},}\ }\href {\doibase 10.1364/OL.469657} {\bibfield
  {journal} {\bibinfo  {journal} {Optics Letters}\ }\textbf {\bibinfo {volume}
  {47}},\ \bibinfo {pages} {4620--4623} (\bibinfo {year} {2022})}\BibitemShut
  {NoStop}%
\bibitem [{\citenamefont {Apigo}\ \emph {et~al.}(2019)\citenamefont {Apigo},
  \citenamefont {Cheng}, \citenamefont {Dobiszewski}, \citenamefont {Prodan},\
  and\ \citenamefont {Prodan}}]{Apigo2019}%
  \BibitemOpen
  \bibfield  {author} {\bibinfo {author} {\bibfnamefont {D.~J.}\ \bibnamefont
  {Apigo}}, \bibinfo {author} {\bibfnamefont {W.}~\bibnamefont {Cheng}},
  \bibinfo {author} {\bibfnamefont {K.~F.}\ \bibnamefont {Dobiszewski}},
  \bibinfo {author} {\bibfnamefont {E.}~\bibnamefont {Prodan}}, \ and\ \bibinfo
  {author} {\bibfnamefont {C.}~\bibnamefont {Prodan}},\ }\bibfield  {title}
  {\enquote {\bibinfo {title} {Observation of topological edge modes in a
  quasiperiodic acoustic waveguide},}\ }\href {\doibase
  10.1103/PhysRevLett.122.095501} {\bibfield  {journal} {\bibinfo  {journal}
  {Phys. Rev. Lett.}\ }\textbf {\bibinfo {volume} {122}},\ \bibinfo {pages}
  {095501} (\bibinfo {year} {2019})}\BibitemShut {NoStop}%
\bibitem [{\citenamefont {Ni}\ \emph {et~al.}(2019)\citenamefont {Ni},
  \citenamefont {Chen}, \citenamefont {Weiner}, \citenamefont {Apigo},
  \citenamefont {Prodan}, \citenamefont {Al\`u}, \citenamefont {Prodan},\ and\
  \citenamefont {Khanikaev}}]{NiX2019Observation}%
  \BibitemOpen
  \bibfield  {author} {\bibinfo {author} {\bibfnamefont {X.}~\bibnamefont
  {Ni}}, \bibinfo {author} {\bibfnamefont {K.}~\bibnamefont {Chen}}, \bibinfo
  {author} {\bibfnamefont {M.}~\bibnamefont {Weiner}}, \bibinfo {author}
  {\bibfnamefont {.~J.}\ \bibnamefont {Apigo}}, \bibinfo {author}
  {\bibfnamefont {C.}~\bibnamefont {Prodan}}, \bibinfo {author} {\bibfnamefont
  {A.}~\bibnamefont {Al\`u}}, \bibinfo {author} {\bibfnamefont
  {E.}~\bibnamefont {Prodan}}, \ and\ \bibinfo {author} {\bibfnamefont {A.~B.}\
  \bibnamefont {Khanikaev}},\ }\bibfield  {title} {\enquote {\bibinfo {title}
  {Observation of {Hofstadter} butterfly and topological edge states in
  reconfigurable quasi-periodic acoustic crystals},}\ }\href {\doibase
  10.1038/s42005-019-0151-7} {\bibfield  {journal} {\bibinfo  {journal}
  {Commun. Phys.}\ }\textbf {\bibinfo {volume} {2}},\ \bibinfo {pages} {55}
  (\bibinfo {year} {2019})}\BibitemShut {NoStop}%
\bibitem [{\citenamefont {Chen}\ \emph {et~al.}(2021)\citenamefont {Chen},
  \citenamefont {Tang}, \citenamefont {Zhang}, \citenamefont {Chen},\ and\
  \citenamefont {Ma}}]{ChenZ2021}%
  \BibitemOpen
  \bibfield  {author} {\bibinfo {author} {\bibfnamefont {Z.-G.}\ \bibnamefont
  {Chen}}, \bibinfo {author} {\bibfnamefont {W.}~\bibnamefont {Tang}}, \bibinfo
  {author} {\bibfnamefont {R.-Y.}\ \bibnamefont {Zhang}}, \bibinfo {author}
  {\bibfnamefont {Z.}~\bibnamefont {Chen}}, \ and\ \bibinfo {author}
  {\bibfnamefont {G.}~\bibnamefont {Ma}},\ }\bibfield  {title} {\enquote
  {\bibinfo {title} {{Landau-Zener} transition in the dynamic transfer of
  acoustic topological states},}\ }\href {\doibase
  10.1103/PhysRevLett.126.054301} {\bibfield  {journal} {\bibinfo  {journal}
  {Phys. Rev. Lett.}\ }\textbf {\bibinfo {volume} {126}} (\bibinfo {year}
  {2021}),\ 10.1103/PhysRevLett.126.054301}\BibitemShut {NoStop}%
\bibitem [{\citenamefont {You}\ \emph {et~al.}(2022)\citenamefont {You},
  \citenamefont {Liang}, \citenamefont {Xie}, \citenamefont {Gao},
  \citenamefont {Ye}, \citenamefont {Zhu},\ and\ \citenamefont
  {Zhang}}]{You2022Observation}%
  \BibitemOpen
  \bibfield  {author} {\bibinfo {author} {\bibfnamefont {O.}~\bibnamefont
  {You}}, \bibinfo {author} {\bibfnamefont {S.}~\bibnamefont {Liang}}, \bibinfo
  {author} {\bibfnamefont {B.}~\bibnamefont {Xie}}, \bibinfo {author}
  {\bibfnamefont {W.}~\bibnamefont {Gao}}, \bibinfo {author} {\bibfnamefont
  {W.}~\bibnamefont {Ye}}, \bibinfo {author} {\bibfnamefont {J.}~\bibnamefont
  {Zhu}}, \ and\ \bibinfo {author} {\bibfnamefont {S.}~\bibnamefont {Zhang}},\
  }\bibfield  {title} {\enquote {\bibinfo {title} {Observation of {Non-Abelian}
  {Thouless} pump},}\ }\href {\doibase 10.1103/PhysRevLett.128.244302}
  {\bibfield  {journal} {\bibinfo  {journal} {Phys. Rev. Lett.}\ }\textbf
  {\bibinfo {volume} {128}},\ \bibinfo {pages} {244302} (\bibinfo {year}
  {2022})}\BibitemShut {NoStop}%
\bibitem [{\citenamefont {Hu}\ \emph {et~al.}(2024)\citenamefont {Hu},
  \citenamefont {Wu}, \citenamefont {Xie}, \citenamefont {Zhuo}, \citenamefont
  {Sun}, \citenamefont {Sheng},\ and\ \citenamefont {Pan}}]{HuP2024Hearing}%
  \BibitemOpen
  \bibfield  {author} {\bibinfo {author} {\bibfnamefont {P.}~\bibnamefont
  {Hu}}, \bibinfo {author} {\bibfnamefont {H.-W.}\ \bibnamefont {Wu}}, \bibinfo
  {author} {\bibfnamefont {P.-X.}\ \bibnamefont {Xie}}, \bibinfo {author}
  {\bibfnamefont {Y.}~\bibnamefont {Zhuo}}, \bibinfo {author} {\bibfnamefont
  {W.}~\bibnamefont {Sun}}, \bibinfo {author} {\bibfnamefont {Z.}~\bibnamefont
  {Sheng}}, \ and\ \bibinfo {author} {\bibfnamefont {Y.}~\bibnamefont {Pan}},\
  }\bibfield  {title} {\enquote {\bibinfo {title} {Hearing the dynamical
  {Floquet-Thouless} pump of a sound pulse},}\ }\href {\doibase
  10.1103/PhysRevB.00.005100} {\bibfield  {journal} {\bibinfo  {journal} {Phys.
  Rev. B}\ }\textbf {\bibinfo {volume} {in print}},\ \bibinfo {pages} {12}
  (\bibinfo {year} {2024})}\BibitemShut {NoStop}%
\bibitem [{\citenamefont {Ke}\ \emph {et~al.}(2016)\citenamefont {Ke},
  \citenamefont {Qin}, \citenamefont {Mei}, \citenamefont {Zhong},
  \citenamefont {Kivshar},\ and\ \citenamefont {Lee}}]{KeY2016}%
  \BibitemOpen
  \bibfield  {author} {\bibinfo {author} {\bibfnamefont {Y.}~\bibnamefont
  {Ke}}, \bibinfo {author} {\bibfnamefont {X.}~\bibnamefont {Qin}}, \bibinfo
  {author} {\bibfnamefont {F.}~\bibnamefont {Mei}}, \bibinfo {author}
  {\bibfnamefont {H.}~\bibnamefont {Zhong}}, \bibinfo {author} {\bibfnamefont
  {Y.~S.}\ \bibnamefont {Kivshar}}, \ and\ \bibinfo {author} {\bibfnamefont
  {C.}~\bibnamefont {Lee}},\ }\bibfield  {title} {\enquote {\bibinfo {title}
  {Topological phase transitions and {Thouless} pumping of light in photonic
  waveguide arrays},}\ }\href {\doibase https://doi.org/10.1002/lpor.201600119}
  {\bibfield  {journal} {\bibinfo  {journal} {Laser \& Photonics Reviews}\
  }\textbf {\bibinfo {volume} {10}},\ \bibinfo {pages} {995--1001} (\bibinfo
  {year} {2016})}\BibitemShut {NoStop}%
\bibitem [{\citenamefont {Meier}\ \emph {et~al.}(2016)\citenamefont {Meier},
  \citenamefont {An},\ and\ \citenamefont {Gadway}}]{Meier2016}%
  \BibitemOpen
  \bibfield  {author} {\bibinfo {author} {\bibfnamefont {E.~J.}\ \bibnamefont
  {Meier}}, \bibinfo {author} {\bibfnamefont {F.~A.}\ \bibnamefont {An}}, \
  and\ \bibinfo {author} {\bibfnamefont {B.}~\bibnamefont {Gadway}},\
  }\bibfield  {title} {\enquote {\bibinfo {title} {Observation of the
  topological soliton state in the {Su–Schrieffer–Heeger} model},}\ }\href
  {\doibase 10.1038/ncomms13986} {\bibfield  {journal} {\bibinfo  {journal}
  {Nat. Commun.}\ }\textbf {\bibinfo {volume} {7}},\ \bibinfo {pages} {13986}
  (\bibinfo {year} {2016})}\BibitemShut {NoStop}%
\bibitem [{\citenamefont {Alpeggiani}\ and\ \citenamefont
  {Kuipers}(2019)}]{Alpeggiani2019Topo}%
  \BibitemOpen
  \bibfield  {author} {\bibinfo {author} {\bibfnamefont {F.}~\bibnamefont
  {Alpeggiani}}\ and\ \bibinfo {author} {\bibfnamefont {L.}~\bibnamefont
  {Kuipers}},\ }\bibfield  {title} {\enquote {\bibinfo {title} {Topological
  edge states in bichromatic photonic crystals},}\ }\href {\doibase
  10.1364/OPTICA.6.000096} {\bibfield  {journal} {\bibinfo  {journal} {Optica}\
  }\textbf {\bibinfo {volume} {6}},\ \bibinfo {pages} {96--103} (\bibinfo
  {year} {2019})}\BibitemShut {NoStop}%
\bibitem [{\citenamefont {Cheng}\ \emph {et~al.}(2022)\citenamefont {Cheng},
  \citenamefont {Wang}, \citenamefont {Ke}, \citenamefont {Chen}, \citenamefont
  {Yu}, \citenamefont {Kivshar}, \citenamefont {Lee},\ and\ \citenamefont
  {Pan}}]{ChengQ2022}%
  \BibitemOpen
  \bibfield  {author} {\bibinfo {author} {\bibfnamefont {Q.}~\bibnamefont
  {Cheng}}, \bibinfo {author} {\bibfnamefont {H.}~\bibnamefont {Wang}},
  \bibinfo {author} {\bibfnamefont {Y.}~\bibnamefont {Ke}}, \bibinfo {author}
  {\bibfnamefont {T.}~\bibnamefont {Chen}}, \bibinfo {author} {\bibfnamefont
  {Y.}~\bibnamefont {Yu}}, \bibinfo {author} {\bibfnamefont {Y.~S.}\
  \bibnamefont {Kivshar}}, \bibinfo {author} {\bibfnamefont {C.}~\bibnamefont
  {Lee}}, \ and\ \bibinfo {author} {\bibfnamefont {Y.}~\bibnamefont {Pan}},\
  }\bibfield  {title} {\enquote {\bibinfo {title} {Asymmetric topological
  pumping in nonparaxial photonics},}\ }\href {\doibase
  10.1038/s41467-021-27773-9} {\bibfield  {journal} {\bibinfo  {journal} {Nat.
  Commun.}\ }\textbf {\bibinfo {volume} {13}},\ \bibinfo {pages} {249}
  (\bibinfo {year} {2022})}\BibitemShut {NoStop}%
\bibitem [{\citenamefont {Adiyatullin}\ \emph {et~al.}(2023)\citenamefont
  {Adiyatullin}, \citenamefont {Upreti}, \citenamefont {Lechevalier},
  \citenamefont {Evain}, \citenamefont {Copie}, \citenamefont {Suret},
  \citenamefont {Randoux}, \citenamefont {Delplace},\ and\ \citenamefont
  {Amo}}]{Adiyatullin2023}%
  \BibitemOpen
  \bibfield  {author} {\bibinfo {author} {\bibfnamefont {A.~F.}\ \bibnamefont
  {Adiyatullin}}, \bibinfo {author} {\bibfnamefont {L.~K.}\ \bibnamefont
  {Upreti}}, \bibinfo {author} {\bibfnamefont {C.}~\bibnamefont {Lechevalier}},
  \bibinfo {author} {\bibfnamefont {C.}~\bibnamefont {Evain}}, \bibinfo
  {author} {\bibfnamefont {F.}~\bibnamefont {Copie}}, \bibinfo {author}
  {\bibfnamefont {P.}~\bibnamefont {Suret}}, \bibinfo {author} {\bibfnamefont
  {S.}~\bibnamefont {Randoux}}, \bibinfo {author} {\bibfnamefont
  {P.}~\bibnamefont {Delplace}}, \ and\ \bibinfo {author} {\bibfnamefont
  {A.}~\bibnamefont {Amo}},\ }\bibfield  {title} {\enquote {\bibinfo {title}
  {Topological properties of {Floquet} winding bands in a photonic lattice},}\
  }\href {\doibase 10.1103/PhysRevLett.130.056901} {\bibfield  {journal}
  {\bibinfo  {journal} {Phys. Rev. Lett.}\ }\textbf {\bibinfo {volume} {130}},\
  \bibinfo {pages} {056901} (\bibinfo {year} {2023})}\BibitemShut {NoStop}%
\bibitem [{\citenamefont {Upreti}\ \emph {et~al.}(2020)\citenamefont {Upreti},
  \citenamefont {Evain}, \citenamefont {Randoux}, \citenamefont {Suret},
  \citenamefont {Amo},\ and\ \citenamefont {Delplace}}]{Upreti2020}%
  \BibitemOpen
  \bibfield  {author} {\bibinfo {author} {\bibfnamefont {L.~K.}\ \bibnamefont
  {Upreti}}, \bibinfo {author} {\bibfnamefont {C.}~\bibnamefont {Evain}},
  \bibinfo {author} {\bibfnamefont {S.}~\bibnamefont {Randoux}}, \bibinfo
  {author} {\bibfnamefont {P.}~\bibnamefont {Suret}}, \bibinfo {author}
  {\bibfnamefont {A.}~\bibnamefont {Amo}}, \ and\ \bibinfo {author}
  {\bibfnamefont {P.}~\bibnamefont {Delplace}},\ }\bibfield  {title} {\enquote
  {\bibinfo {title} {Topological swing of bloch oscillations in quantum
  walks},}\ }\href {\doibase 10.1103/PhysRevLett.125.186804} {\bibfield
  {journal} {\bibinfo  {journal} {Phys. Rev. Lett.}\ }\textbf {\bibinfo
  {volume} {125}},\ \bibinfo {pages} {186804} (\bibinfo {year}
  {2020})}\BibitemShut {NoStop}%
\bibitem [{\citenamefont {Minguzzi}\ \emph {et~al.}(2022)\citenamefont
  {Minguzzi}, \citenamefont {Zhu}, \citenamefont {Sandholzer}, \citenamefont
  {Walter}, \citenamefont {Viebahn},\ and\ \citenamefont
  {Esslinger}}]{Minguzzi2022}%
  \BibitemOpen
  \bibfield  {author} {\bibinfo {author} {\bibfnamefont {J.}~\bibnamefont
  {Minguzzi}}, \bibinfo {author} {\bibfnamefont {Z.}~\bibnamefont {Zhu}},
  \bibinfo {author} {\bibfnamefont {K.}~\bibnamefont {Sandholzer}}, \bibinfo
  {author} {\bibfnamefont {A.-S.}\ \bibnamefont {Walter}}, \bibinfo {author}
  {\bibfnamefont {K.}~\bibnamefont {Viebahn}}, \ and\ \bibinfo {author}
  {\bibfnamefont {T.}~\bibnamefont {Esslinger}},\ }\bibfield  {title} {\enquote
  {\bibinfo {title} {Topological pumping in a {Floquet-Bloch} band},}\ }\href
  {\doibase 10.1103/PhysRevLett.129.053201} {\bibfield  {journal} {\bibinfo
  {journal} {Phys. Rev. Lett.}\ }\textbf {\bibinfo {volume} {129}},\ \bibinfo
  {pages} {053201} (\bibinfo {year} {2022})}\BibitemShut {NoStop}%
\bibitem [{\citenamefont {Schweizer}\ \emph {et~al.}(2016)\citenamefont
  {Schweizer}, \citenamefont {Lohse}, \citenamefont {Citro},\ and\
  \citenamefont {Bloch}}]{Schweizer2016}%
  \BibitemOpen
  \bibfield  {author} {\bibinfo {author} {\bibfnamefont {C.}~\bibnamefont
  {Schweizer}}, \bibinfo {author} {\bibfnamefont {M.}~\bibnamefont {Lohse}},
  \bibinfo {author} {\bibfnamefont {R.}~\bibnamefont {Citro}}, \ and\ \bibinfo
  {author} {\bibfnamefont {I.}~\bibnamefont {Bloch}},\ }\bibfield  {title}
  {\enquote {\bibinfo {title} {Spin pumping and measurement of spin currents in
  optical superlattices},}\ }\href {\doibase 10.1103/PhysRevLett.117.170405}
  {\bibfield  {journal} {\bibinfo  {journal} {Phys. Rev. Lett.}\ }\textbf
  {\bibinfo {volume} {117}} (\bibinfo {year} {2016}),\
  10.1103/PhysRevLett.117.170405}\BibitemShut {NoStop}%
\bibitem [{\citenamefont {Viebahn}\ \emph {et~al.}(2024)\citenamefont
  {Viebahn}, \citenamefont {Walter}, \citenamefont {Bertok}, \citenamefont
  {Zhu}, \citenamefont {G{\"a}chter}, \citenamefont {Aligia}, \citenamefont
  {Heidrich-Meisner},\ and\ \citenamefont
  {Esslinger}}]{Viebahn2024Interactions}%
  \BibitemOpen
  \bibfield  {author} {\bibinfo {author} {\bibfnamefont {K.}~\bibnamefont
  {Viebahn}}, \bibinfo {author} {\bibfnamefont {A.-S.}\ \bibnamefont {Walter}},
  \bibinfo {author} {\bibfnamefont {E.}~\bibnamefont {Bertok}}, \bibinfo
  {author} {\bibfnamefont {Z.}~\bibnamefont {Zhu}}, \bibinfo {author}
  {\bibfnamefont {M.}~\bibnamefont {G{\"a}chter}}, \bibinfo {author}
  {\bibfnamefont {A.~A.}\ \bibnamefont {Aligia}}, \bibinfo {author}
  {\bibfnamefont {F.}~\bibnamefont {Heidrich-Meisner}}, \ and\ \bibinfo
  {author} {\bibfnamefont {T.}~\bibnamefont {Esslinger}},\ }\bibfield  {title}
  {\enquote {\bibinfo {title} {Interactions enable thouless pumping in a
  nonsliding lattice},}\ }\href {\doibase 10.1103/PhysRevX.14.021049}
  {\bibfield  {journal} {\bibinfo  {journal} {Phys. Rev. X}\ }\textbf {\bibinfo
  {volume} {14}},\ \bibinfo {pages} {021049} (\bibinfo {year}
  {2024})}\BibitemShut {NoStop}%
\bibitem [{\citenamefont {Cohen}\ \emph {et~al.}(2019)\citenamefont {Cohen},
  \citenamefont {Larocque}, \citenamefont {Bouchard}, \citenamefont
  {Nejadsattari}, \citenamefont {Gefen},\ and\ \citenamefont
  {Karimi}}]{Cohen2019Geometric}%
  \BibitemOpen
  \bibfield  {author} {\bibinfo {author} {\bibfnamefont {E.}~\bibnamefont
  {Cohen}}, \bibinfo {author} {\bibfnamefont {H.}~\bibnamefont {Larocque}},
  \bibinfo {author} {\bibfnamefont {F.}~\bibnamefont {Bouchard}}, \bibinfo
  {author} {\bibfnamefont {F.}~\bibnamefont {Nejadsattari}}, \bibinfo {author}
  {\bibfnamefont {Y.}~\bibnamefont {Gefen}}, \ and\ \bibinfo {author}
  {\bibfnamefont {E.}~\bibnamefont {Karimi}},\ }\bibfield  {title} {\enquote
  {\bibinfo {title} {Geometric phase from aharonov–bohm to
  pancharatnam–berry and beyond},}\ }\href {\doibase
  https://doi.org/10.1038/s42254-019-0071-1} {\bibfield  {journal} {\bibinfo
  {journal} {{N}at. {R}ev. {P}hys.}\ }\textbf {\bibinfo {volume} {1}},\
  \bibinfo {pages} {437 -- 449} (\bibinfo {year} {2019})}\BibitemShut {NoStop}%
\bibitem [{\citenamefont {Berry}(1984)}]{Berry1984Geometry}%
  \BibitemOpen
  \bibfield  {author} {\bibinfo {author} {\bibfnamefont {M.~V.}\ \bibnamefont
  {Berry}},\ }\bibfield  {title} {\enquote {\bibinfo {title} {Quantal phase
  factors accompanying adiabatic changes},}\ }\href {\doibase
  10.1098/rspa.1984.0023} {\bibfield  {journal} {\bibinfo  {journal}
  {Proceedings of the Royal Society of London. Series A, Mathematical and
  Physical Sciences}\ }\textbf {\bibinfo {volume} {392}},\ \bibinfo {pages}
  {45--57} (\bibinfo {year} {1984})}\BibitemShut {NoStop}%
\bibitem [{\citenamefont {Switkes}\ \emph {et~al.}(1999)\citenamefont
  {Switkes}, \citenamefont {Marcus}, \citenamefont {Campman},\ and\
  \citenamefont {Gossard}}]{Switkes1999Adiabatic}%
  \BibitemOpen
  \bibfield  {author} {\bibinfo {author} {\bibfnamefont {M.}~\bibnamefont
  {Switkes}}, \bibinfo {author} {\bibfnamefont {C.~M.}\ \bibnamefont {Marcus}},
  \bibinfo {author} {\bibfnamefont {K.}~\bibnamefont {Campman}}, \ and\
  \bibinfo {author} {\bibfnamefont {A.~C.}\ \bibnamefont {Gossard}},\
  }\bibfield  {title} {\enquote {\bibinfo {title} {An adiabatic quantum
  electron pump},}\ }\href {\doibase 10.1126/science.283.5409.1905} {\bibfield
  {journal} {\bibinfo  {journal} {Science}\ }\textbf {\bibinfo {volume}
  {283}},\ \bibinfo {pages} {1905--1908} (\bibinfo {year} {1999})}\BibitemShut
  {NoStop}%
\bibitem [{\citenamefont {Verbin}\ \emph {et~al.}(2015)\citenamefont {Verbin},
  \citenamefont {Zilberberg}, \citenamefont {Lahini}, \citenamefont {Kraus},\
  and\ \citenamefont {Silberberg}}]{Verbin2015Topological}%
  \BibitemOpen
  \bibfield  {author} {\bibinfo {author} {\bibfnamefont {M.}~\bibnamefont
  {Verbin}}, \bibinfo {author} {\bibfnamefont {O.}~\bibnamefont {Zilberberg}},
  \bibinfo {author} {\bibfnamefont {Y.}~\bibnamefont {Lahini}}, \bibinfo
  {author} {\bibfnamefont {Y.~E.}\ \bibnamefont {Kraus}}, \ and\ \bibinfo
  {author} {\bibfnamefont {Y.}~\bibnamefont {Silberberg}},\ }\bibfield  {title}
  {\enquote {\bibinfo {title} {Topological pumping over a photonic {Fibonacci}
  quasicrystal},}\ }\href {\doibase 10.1103/PhysRevB.91.064201} {\bibfield
  {journal} {\bibinfo  {journal} {Phys. Rev. B}\ }\textbf {\bibinfo {volume}
  {91}},\ \bibinfo {pages} {064201} (\bibinfo {year} {2015})}\BibitemShut
  {NoStop}%
\bibitem [{\citenamefont {Verbin}\ \emph {et~al.}(2013)\citenamefont {Verbin},
  \citenamefont {Zilberberg}, \citenamefont {Kraus}, \citenamefont {Lahini},\
  and\ \citenamefont {Silberberg}}]{Verbin2013}%
  \BibitemOpen
  \bibfield  {author} {\bibinfo {author} {\bibfnamefont {M.}~\bibnamefont
  {Verbin}}, \bibinfo {author} {\bibfnamefont {O.}~\bibnamefont {Zilberberg}},
  \bibinfo {author} {\bibfnamefont {Y.~E.}\ \bibnamefont {Kraus}}, \bibinfo
  {author} {\bibfnamefont {Y.}~\bibnamefont {Lahini}}, \ and\ \bibinfo {author}
  {\bibfnamefont {Y.}~\bibnamefont {Silberberg}},\ }\bibfield  {title}
  {\enquote {\bibinfo {title} {Observation of topological phase transitions in
  photonic quasicrystals},}\ }\href {\doibase 10.1103/PhysRevLett.110.076403}
  {\bibfield  {journal} {\bibinfo  {journal} {Phys. Rev. Lett.}\ }\textbf
  {\bibinfo {volume} {110}},\ \bibinfo {pages} {076403} (\bibinfo {year}
  {2013})}\BibitemShut {NoStop}%
\bibitem [{\citenamefont {Xiang}\ \emph {et~al.}(2023)\citenamefont {Xiang},
  \citenamefont {Huang}, \citenamefont {Zhang}, \citenamefont {Liu},
  \citenamefont {Shi}, \citenamefont {Deng}, \citenamefont {Liu}, \citenamefont
  {Li}, \citenamefont {Liang}, \citenamefont {Mei}, \citenamefont {Yu},
  \citenamefont {Xue}, \citenamefont {Tian}, \citenamefont {Song},
  \citenamefont {Liu}, \citenamefont {Xu}, \citenamefont {Zheng}, \citenamefont
  {Nori},\ and\ \citenamefont {Fan}}]{XiangZ2023Simulating}%
  \BibitemOpen
  \bibfield  {author} {\bibinfo {author} {\bibfnamefont {Z.~C.}\ \bibnamefont
  {Xiang}}, \bibinfo {author} {\bibfnamefont {K.}~\bibnamefont {Huang}},
  \bibinfo {author} {\bibfnamefont {Y.~R.}\ \bibnamefont {Zhang}}, \bibinfo
  {author} {\bibfnamefont {T.}~\bibnamefont {Liu}}, \bibinfo {author}
  {\bibfnamefont {Y.~H.}\ \bibnamefont {Shi}}, \bibinfo {author} {\bibfnamefont
  {C.~L.}\ \bibnamefont {Deng}}, \bibinfo {author} {\bibfnamefont
  {T.}~\bibnamefont {Liu}}, \bibinfo {author} {\bibfnamefont {H.}~\bibnamefont
  {Li}}, \bibinfo {author} {\bibfnamefont {G.~H.}\ \bibnamefont {Liang}},
  \bibinfo {author} {\bibfnamefont {Z.~Y.}\ \bibnamefont {Mei}}, \bibinfo
  {author} {\bibfnamefont {H.}~\bibnamefont {Yu}}, \bibinfo {author}
  {\bibfnamefont {G.}~\bibnamefont {Xue}}, \bibinfo {author} {\bibfnamefont
  {Y.}~\bibnamefont {Tian}}, \bibinfo {author} {\bibfnamefont {X.}~\bibnamefont
  {Song}}, \bibinfo {author} {\bibfnamefont {Z.~B.}\ \bibnamefont {Liu}},
  \bibinfo {author} {\bibfnamefont {K.}~\bibnamefont {Xu}}, \bibinfo {author}
  {\bibfnamefont {D.}~\bibnamefont {Zheng}}, \bibinfo {author} {\bibfnamefont
  {F.}~\bibnamefont {Nori}}, \ and\ \bibinfo {author} {\bibfnamefont
  {H.}~\bibnamefont {Fan}},\ }\bibfield  {title} {\enquote {\bibinfo {title}
  {Simulating {Chern} insulators on a superconducting quantum processor},}\
  }\href {\doibase 10.1038/s41467-023-41230-9} {\bibfield  {journal} {\bibinfo
  {journal} {Nat. Commun.}\ }\textbf {\bibinfo {volume} {14}},\ \bibinfo
  {pages} {5433} (\bibinfo {year} {2023})}\BibitemShut {NoStop}%
\bibitem [{\citenamefont {Rajagopal}\ \emph {et~al.}(2019)\citenamefont
  {Rajagopal}, \citenamefont {Shimasaki}, \citenamefont {Dotti}, \citenamefont
  {Raciunas}, \citenamefont {Senaratne}, \citenamefont {Anisimovas},
  \citenamefont {Eckardt},\ and\ \citenamefont {Weld}}]{Rajagopal2019}%
  \BibitemOpen
  \bibfield  {author} {\bibinfo {author} {\bibfnamefont {S.~V.}\ \bibnamefont
  {Rajagopal}}, \bibinfo {author} {\bibfnamefont {T.}~\bibnamefont
  {Shimasaki}}, \bibinfo {author} {\bibfnamefont {P.}~\bibnamefont {Dotti}},
  \bibinfo {author} {\bibfnamefont {M.}~\bibnamefont {Raciunas}}, \bibinfo
  {author} {\bibfnamefont {R.}~\bibnamefont {Senaratne}}, \bibinfo {author}
  {\bibfnamefont {E.}~\bibnamefont {Anisimovas}}, \bibinfo {author}
  {\bibfnamefont {A.}~\bibnamefont {Eckardt}}, \ and\ \bibinfo {author}
  {\bibfnamefont {D.~M.}\ \bibnamefont {Weld}},\ }\bibfield  {title} {\enquote
  {\bibinfo {title} {Phasonic spectroscopy of a quantum gas in a
  quasicrystalline lattice},}\ }\href {\doibase 10.1103/PhysRevLett.123.223201}
  {\bibfield  {journal} {\bibinfo  {journal} {Phys. Rev. Lett.}\ }\textbf
  {\bibinfo {volume} {123}},\ \bibinfo {pages} {223201} (\bibinfo {year}
  {2019})}\BibitemShut {NoStop}%
\bibitem [{\citenamefont {Yu}\ \emph {et~al.}(2020)\citenamefont {Yu},
  \citenamefont {Qiu}, \citenamefont {Chong}, \citenamefont {Torquato},\ and\
  \citenamefont {Park}}]{YuS2020}%
  \BibitemOpen
  \bibfield  {author} {\bibinfo {author} {\bibfnamefont {S.}~\bibnamefont
  {Yu}}, \bibinfo {author} {\bibfnamefont {C.-W.}\ \bibnamefont {Qiu}},
  \bibinfo {author} {\bibfnamefont {Y.}~\bibnamefont {Chong}}, \bibinfo
  {author} {\bibfnamefont {S.}~\bibnamefont {Torquato}}, \ and\ \bibinfo
  {author} {\bibfnamefont {N.}~\bibnamefont {Park}},\ }\bibfield  {title}
  {\enquote {\bibinfo {title} {Engineered disorder in photonics},}\ }\href
  {\doibase 10.1038/s41578-020-00263-y} {\bibfield  {journal} {\bibinfo
  {journal} {Nat. Rev. Mater.}\ }\textbf {\bibinfo {volume} {6}},\ \bibinfo
  {pages} {226--243} (\bibinfo {year} {2020})}\BibitemShut {NoStop}%
\bibitem [{\citenamefont {Berry}(2009)}]{Berry2009Transitionless}%
  \BibitemOpen
  \bibfield  {author} {\bibinfo {author} {\bibfnamefont {M.~V.}\ \bibnamefont
  {Berry}},\ }\bibfield  {title} {\enquote {\bibinfo {title} {Transitionless
  quantum driving},}\ }\href {\doibase 10.1088/1751-8113/42/36/365303}
  {\bibfield  {journal} {\bibinfo  {journal} {Journal of Physics A:
  Mathematical and Theoretical}\ }\textbf {\bibinfo {volume} {42}},\ \bibinfo
  {pages} {365303} (\bibinfo {year} {2009})}\BibitemShut {NoStop}%
\bibitem [{\citenamefont {Shevchenko}\ \emph {et~al.}(2010)\citenamefont
  {Shevchenko}, \citenamefont {Ashhab},\ and\ \citenamefont
  {Nori}}]{Shevchenko2010LZS}%
  \BibitemOpen
  \bibfield  {author} {\bibinfo {author} {\bibfnamefont {S.~N.}\ \bibnamefont
  {Shevchenko}}, \bibinfo {author} {\bibfnamefont {S.}~\bibnamefont {Ashhab}},
  \ and\ \bibinfo {author} {\bibfnamefont {F.}~\bibnamefont {Nori}},\
  }\bibfield  {title} {\enquote {\bibinfo {title}
  {{Landau-Zener-St{\"u}ckelberg} interferometry},}\ }\href {\doibase
  https://doi.org/10.1016/j.physrep.2010.03.002} {\bibfield  {journal}
  {\bibinfo  {journal} {Physics Reports}\ }\textbf {\bibinfo {volume} {492}},\
  \bibinfo {pages} {1--30} (\bibinfo {year} {2010})}\BibitemShut {NoStop}%
\bibitem [{\citenamefont {Wang}\ \emph {et~al.}(2015)\citenamefont {Wang},
  \citenamefont {Zhang}, \citenamefont {Zhang}, \citenamefont {Raithel},
  \citenamefont {Zhao},\ and\ \citenamefont
  {Jia}}]{WangL2015Atom-interferometric}%
  \BibitemOpen
  \bibfield  {author} {\bibinfo {author} {\bibfnamefont {L.}~\bibnamefont
  {Wang}}, \bibinfo {author} {\bibfnamefont {H.}~\bibnamefont {Zhang}},
  \bibinfo {author} {\bibfnamefont {L.}~\bibnamefont {Zhang}}, \bibinfo
  {author} {\bibfnamefont {G.}~\bibnamefont {Raithel}}, \bibinfo {author}
  {\bibfnamefont {J.}~\bibnamefont {Zhao}}, \ and\ \bibinfo {author}
  {\bibfnamefont {S.}~\bibnamefont {Jia}},\ }\bibfield  {title} {\enquote
  {\bibinfo {title} {Atom-interferometric measurement of {Stark} level
  splittings},}\ }\href {\doibase 10.1103/PhysRevA.92.033619} {\bibfield
  {journal} {\bibinfo  {journal} {Phys. Rev. A}\ }\textbf {\bibinfo {volume}
  {92}},\ \bibinfo {pages} {033619} (\bibinfo {year} {2015})}\BibitemShut
  {NoStop}%
\bibitem [{\citenamefont {Boyers}\ \emph {et~al.}(2020)\citenamefont {Boyers},
  \citenamefont {Crowley}, \citenamefont {Chandran},\ and\ \citenamefont
  {Sushkov}}]{Boyers2020}%
  \BibitemOpen
  \bibfield  {author} {\bibinfo {author} {\bibfnamefont {E.}~\bibnamefont
  {Boyers}}, \bibinfo {author} {\bibfnamefont {P.~J.~D.}\ \bibnamefont
  {Crowley}}, \bibinfo {author} {\bibfnamefont {A.}~\bibnamefont {Chandran}}, \
  and\ \bibinfo {author} {\bibfnamefont {A.~O.}\ \bibnamefont {Sushkov}},\
  }\bibfield  {title} {\enquote {\bibinfo {title} {Exploring 2d synthetic
  quantum {Hall} physics with a quasiperiodically driven qubit},}\ }\href
  {\doibase 10.1103/PhysRevLett.125.160505} {\bibfield  {journal} {\bibinfo
  {journal} {Phys. Rev. Lett.}\ }\textbf {\bibinfo {volume} {125}},\ \bibinfo
  {pages} {160505} (\bibinfo {year} {2020})}\BibitemShut {NoStop}%
\bibitem [{\citenamefont {Qiao}\ \emph {et~al.}(2023)\citenamefont {Qiao},
  \citenamefont {Zhang}, \citenamefont {Jian}, \citenamefont {Ma},
  \citenamefont {Gao}, \citenamefont {Zhang},\ and\ \citenamefont
  {Xue}}]{QiaoX2023Nonlinear}%
  \BibitemOpen
  \bibfield  {author} {\bibinfo {author} {\bibfnamefont {X.}~\bibnamefont
  {Qiao}}, \bibinfo {author} {\bibfnamefont {X.~B.}\ \bibnamefont {Zhang}},
  \bibinfo {author} {\bibfnamefont {Y.}~\bibnamefont {Jian}}, \bibinfo {author}
  {\bibfnamefont {Y.~E.}\ \bibnamefont {Ma}}, \bibinfo {author} {\bibfnamefont
  {R.}~\bibnamefont {Gao}}, \bibinfo {author} {\bibfnamefont {A.~X.}\
  \bibnamefont {Zhang}}, \ and\ \bibinfo {author} {\bibfnamefont {J.~K.}\
  \bibnamefont {Xue}},\ }\bibfield  {title} {\enquote {\bibinfo {title}
  {Nonlinear {Landau-Zener-Stuckelberg-Majorana} tunneling and interferometry
  of extended {Bose-Hubbard} flux ladders},}\ }\href {\doibase
  10.1103/PhysRevE.108.034214} {\bibfield  {journal} {\bibinfo  {journal}
  {Phys. Rev. E}\ }\textbf {\bibinfo {volume} {108}},\ \bibinfo {pages}
  {034214} (\bibinfo {year} {2023})}\BibitemShut {NoStop}%
\bibitem [{\citenamefont {Schomerus}(2013)}]{Schomerus2013Topo}%
  \BibitemOpen
  \bibfield  {author} {\bibinfo {author} {\bibfnamefont {H.}~\bibnamefont
  {Schomerus}},\ }\bibfield  {title} {\enquote {\bibinfo {title} {Topologically
  protected midgap states in complex photonic lattices},}\ }\href {\doibase
  10.1364/OL.38.001912} {\bibfield  {journal} {\bibinfo  {journal} {Opt.
  Lett.}\ }\textbf {\bibinfo {volume} {38}},\ \bibinfo {pages} {1912--4}
  (\bibinfo {year} {2013})}\BibitemShut {NoStop}%
\bibitem [{\citenamefont {Viebahn}\ \emph {et~al.}(2019)\citenamefont
  {Viebahn}, \citenamefont {Sbroscia}, \citenamefont {Carter}, \citenamefont
  {Yu},\ and\ \citenamefont {Schneider}}]{Viebahn2019}%
  \BibitemOpen
  \bibfield  {author} {\bibinfo {author} {\bibfnamefont {K.}~\bibnamefont
  {Viebahn}}, \bibinfo {author} {\bibfnamefont {M.}~\bibnamefont {Sbroscia}},
  \bibinfo {author} {\bibfnamefont {E.}~\bibnamefont {Carter}}, \bibinfo
  {author} {\bibfnamefont {Jr-Chiun}\ \bibnamefont {Yu}}, \ and\ \bibinfo
  {author} {\bibfnamefont {U.}~\bibnamefont {Schneider}},\ }\bibfield  {title}
  {\enquote {\bibinfo {title} {Matter-wave diffraction from a quasicrystalline
  optical lattice},}\ }\href {\doibase 10.1103/PhysRevLett.122.110404}
  {\bibfield  {journal} {\bibinfo  {journal} {Phys. Rev. Lett.}\ }\textbf
  {\bibinfo {volume} {122}},\ \bibinfo {pages} {110404} (\bibinfo {year}
  {2019})}\BibitemShut {NoStop}%
\bibitem [{\citenamefont {Li}\ \emph {et~al.}(2023)\citenamefont {Li},
  \citenamefont {Wang}, \citenamefont {Shi}, \citenamefont {Huang},
  \citenamefont {Song}, \citenamefont {Liang}, \citenamefont {Mei},
  \citenamefont {Zhou}, \citenamefont {Zhang}, \citenamefont {Zhang},
  \citenamefont {Chen}, \citenamefont {Zhao}, \citenamefont {Tian},
  \citenamefont {Yang}, \citenamefont {Xiang}, \citenamefont {Xu},
  \citenamefont {Zheng},\ and\ \citenamefont {Fan}}]{Li2023Observation}%
  \BibitemOpen
  \bibfield  {author} {\bibinfo {author} {\bibfnamefont {H.}~\bibnamefont
  {Li}}, \bibinfo {author} {\bibfnamefont {Y.-Y.}\ \bibnamefont {Wang}},
  \bibinfo {author} {\bibfnamefont {Y.-H.~Shi}\ \bibnamefont {Shi}}, \bibinfo
  {author} {\bibfnamefont {K.}~\bibnamefont {Huang}}, \bibinfo {author}
  {\bibfnamefont {X.}~\bibnamefont {Song}}, \bibinfo {author} {\bibfnamefont
  {G.-H.}\ \bibnamefont {Liang}}, \bibinfo {author} {\bibfnamefont {Z.-Y.}\
  \bibnamefont {Mei}}, \bibinfo {author} {\bibfnamefont {B.}~\bibnamefont
  {Zhou}}, \bibinfo {author} {\bibfnamefont {H.}~\bibnamefont {Zhang}},
  \bibinfo {author} {\bibfnamefont {J.-C.}\ \bibnamefont {Zhang}}, \bibinfo
  {author} {\bibfnamefont {S.}~\bibnamefont {Chen}}, \bibinfo {author}
  {\bibfnamefont {S.~P.}\ \bibnamefont {Zhao}}, \bibinfo {author}
  {\bibfnamefont {Y.}~\bibnamefont {Tian}}, \bibinfo {author} {\bibfnamefont
  {Z.-Y.}\ \bibnamefont {Yang}}, \bibinfo {author} {\bibfnamefont
  {Z.}~\bibnamefont {Xiang}}, \bibinfo {author} {\bibfnamefont
  {K.}~\bibnamefont {Xu}}, \bibinfo {author} {\bibfnamefont {D.}~\bibnamefont
  {Zheng}}, \ and\ \bibinfo {author} {\bibfnamefont {H.}~\bibnamefont {Fan}},\
  }\bibfield  {title} {\enquote {\bibinfo {title} {Observation of critical
  phase transition in a generalized {Aubry-Andr\'e-Harper} model with
  superconducting circuits},}\ }\href {\doibase 10.1038/s41534-023-00712-w}
  {\bibfield  {journal} {\bibinfo  {journal} {{NPJ} Quantum Inf.}\ }\textbf
  {\bibinfo {volume} {9}},\ \bibinfo {pages} {40} (\bibinfo {year}
  {2023})}\BibitemShut {NoStop}%
\bibitem [{\citenamefont {Mei}\ \emph {et~al.}(2018)\citenamefont {Mei},
  \citenamefont {Chen}, \citenamefont {Tian}, \citenamefont {Zhu},\ and\
  \citenamefont {Jia}}]{Mei2018Robust}%
  \BibitemOpen
  \bibfield  {author} {\bibinfo {author} {\bibfnamefont {F.}~\bibnamefont
  {Mei}}, \bibinfo {author} {\bibfnamefont {G.}~\bibnamefont {Chen}}, \bibinfo
  {author} {\bibfnamefont {L.}~\bibnamefont {Tian}}, \bibinfo {author}
  {\bibfnamefont {S.-L.}\ \bibnamefont {Zhu}}, \ and\ \bibinfo {author}
  {\bibfnamefont {S.}~\bibnamefont {Jia}},\ }\bibfield  {title} {\enquote
  {\bibinfo {title} {Robust quantum state transfer via topological edge states
  in superconducting qubit chains},}\ }\href {\doibase
  10.1103/PhysRevA.98.012331} {\bibfield  {journal} {\bibinfo  {journal} {Phys.
  Rev. A}\ }\textbf {\bibinfo {volume} {98}},\ \bibinfo {pages} {012331}
  (\bibinfo {year} {2018})}\BibitemShut {NoStop}%
\bibitem [{\citenamefont {Aubry}\ and\ \citenamefont
  {Andr\'e}(1980)}]{Aubry1980Analyticity}%
  \BibitemOpen
  \bibfield  {author} {\bibinfo {author} {\bibfnamefont {S.}~\bibnamefont
  {Aubry}}\ and\ \bibinfo {author} {\bibfnamefont {G.}~\bibnamefont
  {Andr\'e}},\ }\bibfield  {title} {\enquote {\bibinfo {title} {Analyticity
  breaking and {Anderson} localization in incommensurate lattices},}\
  }\href@noop {} {\bibfield  {journal} {\bibinfo  {journal} {Annals of the
  Israel Phyical Society}\ }\textbf {\bibinfo {volume} {3}} (\bibinfo {year}
  {1980})}\BibitemShut {NoStop}%
\bibitem [{\citenamefont {Ganeshan}\ \emph {et~al.}(2013)\citenamefont
  {Ganeshan}, \citenamefont {Sun},\ and\ \citenamefont
  {Das~Sarma}}]{Ganeshan2013Topological}%
  \BibitemOpen
  \bibfield  {author} {\bibinfo {author} {\bibfnamefont {S.}~\bibnamefont
  {Ganeshan}}, \bibinfo {author} {\bibfnamefont {K.}~\bibnamefont {Sun}}, \
  and\ \bibinfo {author} {\bibfnamefont {S.}~\bibnamefont {Das~Sarma}},\
  }\bibfield  {title} {\enquote {\bibinfo {title} {Topological zero-energy
  modes in gapless commensurate {Aubry-Andr\'e-Harper} models},}\ }\href
  {\doibase 10.1103/PhysRevLett.110.180403} {\bibfield  {journal} {\bibinfo
  {journal} {Phys. Rev. Lett.}\ }\textbf {\bibinfo {volume} {110}},\ \bibinfo
  {pages} {180403} (\bibinfo {year} {2013})}\BibitemShut {NoStop}%
\bibitem [{\citenamefont {Yahyavi}\ \emph {et~al.}(2019)\citenamefont
  {Yahyavi}, \citenamefont {Het\'enyi},\ and\ \citenamefont
  {Tanatar}}]{Yahyavi2019Generalized}%
  \BibitemOpen
  \bibfield  {author} {\bibinfo {author} {\bibfnamefont {M.}~\bibnamefont
  {Yahyavi}}, \bibinfo {author} {\bibfnamefont {B.}~\bibnamefont {Het\'enyi}},
  \ and\ \bibinfo {author} {\bibfnamefont {B.}~\bibnamefont {Tanatar}},\
  }\bibfield  {title} {\enquote {\bibinfo {title} {Generalized
  {Aubry-Andr\'e-Harper} model with modulated hopping and $p$-wave pairing},}\
  }\href {\doibase 10.1103/PhysRevB.100.064202} {\bibfield  {journal} {\bibinfo
   {journal} {Phys. Rev. B}\ }\textbf {\bibinfo {volume} {100}},\ \bibinfo
  {pages} {064202} (\bibinfo {year} {2019})}\BibitemShut {NoStop}%
\bibitem [{\citenamefont {Liu}\ and\ \citenamefont
  {Guo}(2018)}]{Liu2018Mobility}%
  \BibitemOpen
  \bibfield  {author} {\bibinfo {author} {\bibfnamefont {T.}~\bibnamefont
  {Liu}}\ and\ \bibinfo {author} {\bibfnamefont {H.}~\bibnamefont {Guo}},\
  }\bibfield  {title} {\enquote {\bibinfo {title} {Mobility edges in
  off-diagonal disordered tight-binding models},}\ }\href {\doibase
  10.1103/PhysRevB.98.104201} {\bibfield  {journal} {\bibinfo  {journal} {Phys.
  Rev. B}\ }\textbf {\bibinfo {volume} {98}},\ \bibinfo {pages} {104201}
  (\bibinfo {year} {2018})}\BibitemShut {NoStop}%
\bibitem [{\citenamefont {Grabarits}\ \emph {et~al.}(2024)\citenamefont
  {Grabarits}, \citenamefont {Tak\'acs}, \citenamefont {Fulga},\ and\
  \citenamefont {Asb\'oth}}]{Grabarits2024Floquet-Anderson}%
  \BibitemOpen
  \bibfield  {author} {\bibinfo {author} {\bibfnamefont {A.}~\bibnamefont
  {Grabarits}}, \bibinfo {author} {\bibfnamefont {A.}~\bibnamefont {Tak\'acs}},
  \bibinfo {author} {\bibfnamefont {I.~C.}\ \bibnamefont {Fulga}}, \ and\
  \bibinfo {author} {\bibfnamefont {J.~K.}\ \bibnamefont {Asb\'oth}},\
  }\bibfield  {title} {\enquote {\bibinfo {title} {{Floquet-Anderson}
  localization in the {Thouless} pump and how to avoid it},}\ }\href {\doibase
  10.1103/PhysRevB.109.L180202} {\bibfield  {journal} {\bibinfo  {journal}
  {Phys. Rev. B}\ }\textbf {\bibinfo {volume} {109}} (\bibinfo {year} {2024}),\
  10.1103/PhysRevB.109.L180202}\BibitemShut {NoStop}%
\bibitem [{\citenamefont {Huang}\ \emph {et~al.}(2011)\citenamefont {Huang},
  \citenamefont {Zhou}, \citenamefont {Fang}, \citenamefont {Kong},
  \citenamefont {Xu}, \citenamefont {Ju},\ and\ \citenamefont
  {Du}}]{Huang2011LZS}%
  \BibitemOpen
  \bibfield  {author} {\bibinfo {author} {\bibfnamefont {P.}~\bibnamefont
  {Huang}}, \bibinfo {author} {\bibfnamefont {J.}~\bibnamefont {Zhou}},
  \bibinfo {author} {\bibfnamefont {F.}~\bibnamefont {Fang}}, \bibinfo {author}
  {\bibfnamefont {X.}~\bibnamefont {Kong}}, \bibinfo {author} {\bibfnamefont
  {X.}~\bibnamefont {Xu}}, \bibinfo {author} {\bibfnamefont {C.}~\bibnamefont
  {Ju}}, \ and\ \bibinfo {author} {\bibfnamefont {J.}~\bibnamefont {Du}},\
  }\bibfield  {title} {\enquote {\bibinfo {title}
  {{Landau-Zener-St{\"u}ckelberg} interferometry of a single electronic spin in
  a noisy environment},}\ }\href {\doibase 10.1103/PhysRevX.1.011003}
  {\bibfield  {journal} {\bibinfo  {journal} {Phys. Rev. X}\ }\textbf {\bibinfo
  {volume} {1}},\ \bibinfo {pages} {011003} (\bibinfo {year}
  {2011})}\BibitemShut {NoStop}%
\bibitem [{\citenamefont {Quintana}\ \emph {et~al.}(2013)\citenamefont
  {Quintana}, \citenamefont {Petersson}, \citenamefont {McFaul}, \citenamefont
  {Srinivasan}, \citenamefont {Houck},\ and\ \citenamefont
  {Petta}}]{Quintana2013Cavity}%
  \BibitemOpen
  \bibfield  {author} {\bibinfo {author} {\bibfnamefont {C.~M.}\ \bibnamefont
  {Quintana}}, \bibinfo {author} {\bibfnamefont {K.~D.}\ \bibnamefont
  {Petersson}}, \bibinfo {author} {\bibfnamefont {L.~W.}\ \bibnamefont
  {McFaul}}, \bibinfo {author} {\bibfnamefont {S.~J.}\ \bibnamefont
  {Srinivasan}}, \bibinfo {author} {\bibfnamefont {A.~A.}\ \bibnamefont
  {Houck}}, \ and\ \bibinfo {author} {\bibfnamefont {J.~R.}\ \bibnamefont
  {Petta}},\ }\bibfield  {title} {\enquote {\bibinfo {title} {Cavity-mediated
  entanglement generation via {Landau-Zener} interferometry},}\ }\href
  {\doibase 10.1103/PhysRevLett.110.173603} {\bibfield  {journal} {\bibinfo
  {journal} {Phys. Rev. Lett.}\ }\textbf {\bibinfo {volume} {110}},\ \bibinfo
  {pages} {173603} (\bibinfo {year} {2013})}\BibitemShut {NoStop}%
\bibitem [{\citenamefont {Korkusinski}\ \emph {et~al.}(2017)\citenamefont
  {Korkusinski}, \citenamefont {Studenikin}, \citenamefont {Aers},
  \citenamefont {Granger}, \citenamefont {Kam},\ and\ \citenamefont
  {Sachrajda}}]{Korkusinski2017LZS}%
  \BibitemOpen
  \bibfield  {author} {\bibinfo {author} {\bibfnamefont {M.}~\bibnamefont
  {Korkusinski}}, \bibinfo {author} {\bibfnamefont {S.~A.}\ \bibnamefont
  {Studenikin}}, \bibinfo {author} {\bibfnamefont {G.}~\bibnamefont {Aers}},
  \bibinfo {author} {\bibfnamefont {G.}~\bibnamefont {Granger}}, \bibinfo
  {author} {\bibfnamefont {A.}~\bibnamefont {Kam}}, \ and\ \bibinfo {author}
  {\bibfnamefont {A.~S.}\ \bibnamefont {Sachrajda}},\ }\bibfield  {title}
  {\enquote {\bibinfo {title} {{L}andau-{Z}ener-{S}t{\"u}ckelberg
  interferometry in quantum dots with fast rise times: Evidence for coherent
  phonon driving},}\ }\href {\doibase 10.1103/PhysRevLett.118.067701}
  {\bibfield  {journal} {\bibinfo  {journal} {Phys. Rev. Lett.}\ }\textbf
  {\bibinfo {volume} {118}},\ \bibinfo {pages} {067701} (\bibinfo {year}
  {2017})}\BibitemShut {NoStop}%
\bibitem [{\citenamefont {Wang}\ \emph
  {et~al.}(2023{\natexlab{b}})\citenamefont {Wang}, \citenamefont {Qin},
  \citenamefont {Zhao}, \citenamefont {Ye}, \citenamefont {Longhi},
  \citenamefont {Lu},\ and\ \citenamefont {Wang}}]{WangS2023Photonic}%
  \BibitemOpen
  \bibfield  {author} {\bibinfo {author} {\bibfnamefont {Shulin}\ \bibnamefont
  {Wang}}, \bibinfo {author} {\bibfnamefont {Chengzhi}\ \bibnamefont {Qin}},
  \bibinfo {author} {\bibfnamefont {Lange}\ \bibnamefont {Zhao}}, \bibinfo
  {author} {\bibfnamefont {Han}\ \bibnamefont {Ye}}, \bibinfo {author}
  {\bibfnamefont {Stefano}\ \bibnamefont {Longhi}}, \bibinfo {author}
  {\bibfnamefont {Peixiang}\ \bibnamefont {Lu}}, \ and\ \bibinfo {author}
  {\bibfnamefont {Bing}\ \bibnamefont {Wang}},\ }\bibfield  {title} {\enquote
  {\bibinfo {title} {Photonic {Floquet Landau-Zener} tunneling and temporal
  beam splitters},}\ }\href {\doibase 10.1126/sciadv.adh0415} {\bibfield
  {journal} {\bibinfo  {journal} {Science Advances}\ }\textbf {\bibinfo
  {volume} {9}},\ \bibinfo {pages} {eadh0415} (\bibinfo {year}
  {2023}{\natexlab{b}})}\BibitemShut {NoStop}%
\bibitem [{\citenamefont {Longhi}(2019)}]{Longhi2019Topo}%
  \BibitemOpen
  \bibfield  {author} {\bibinfo {author} {\bibfnamefont {S.}~\bibnamefont
  {Longhi}},\ }\bibfield  {title} {\enquote {\bibinfo {title} {Topological
  pumping of edge states via adiabatic passage},}\ }\href {\doibase
  10.1103/PhysRevB.99.155150} {\bibfield  {journal} {\bibinfo  {journal} {Phys.
  Rev. B}\ }\textbf {\bibinfo {volume} {99}},\ \bibinfo {pages} {155150}
  (\bibinfo {year} {2019})}\BibitemShut {NoStop}%
\bibitem [{\citenamefont {Aharonov}\ and\ \citenamefont
  {Anandan}(1987)}]{Aharonov1987Phase}%
  \BibitemOpen
  \bibfield  {author} {\bibinfo {author} {\bibfnamefont {Y.}~\bibnamefont
  {Aharonov}}\ and\ \bibinfo {author} {\bibfnamefont {J.}~\bibnamefont
  {Anandan}},\ }\bibfield  {title} {\enquote {\bibinfo {title} {Phase change
  during a cyclic quantum evolution},}\ }\href {\doibase
  10.1103/PhysRevLett.58.1593} {\bibfield  {journal} {\bibinfo  {journal}
  {Phys. Rev. Lett.}\ }\textbf {\bibinfo {volume} {58}},\ \bibinfo {pages}
  {1593--1596} (\bibinfo {year} {1987})}\BibitemShut {NoStop}%
\bibitem [{\citenamefont {Amin}(2009)}]{Amin2009}%
  \BibitemOpen
  \bibfield  {author} {\bibinfo {author} {\bibfnamefont {M.~H.}\ \bibnamefont
  {Amin}},\ }\bibfield  {title} {\enquote {\bibinfo {title} {Consistency of the
  adiabatic theorem},}\ }\href {\doibase 10.1103/PhysRevLett.102.220401}
  {\bibfield  {journal} {\bibinfo  {journal} {Phys. Rev. Lett.}\ }\textbf
  {\bibinfo {volume} {102}},\ \bibinfo {pages} {220401} (\bibinfo {year}
  {2009})}\BibitemShut {NoStop}%
\bibitem [{\citenamefont {Marzlin}\ and\ \citenamefont
  {Sanders}(2004)}]{Marzlin2004Inconsitency}%
  \BibitemOpen
  \bibfield  {author} {\bibinfo {author} {\bibfnamefont {K.~P.}\ \bibnamefont
  {Marzlin}}\ and\ \bibinfo {author} {\bibfnamefont {B.~C.}\ \bibnamefont
  {Sanders}},\ }\bibfield  {title} {\enquote {\bibinfo {title} {Inconsistency
  in the application of the adiabatic theorem},}\ }\href {\doibase
  10.1103/PhysRevLett.93.160408} {\bibfield  {journal} {\bibinfo  {journal}
  {Phys. Rev. Lett.}\ }\textbf {\bibinfo {volume} {93}},\ \bibinfo {pages}
  {160408} (\bibinfo {year} {2004})}\BibitemShut {NoStop}%
\bibitem [{\citenamefont {Riva}\ \emph {et~al.}(2021)\citenamefont {Riva},
  \citenamefont {Castaldini},\ and\ \citenamefont
  {Braghin}}]{Riva2021Adiabatic}%
  \BibitemOpen
  \bibfield  {author} {\bibinfo {author} {\bibfnamefont {E.}~\bibnamefont
  {Riva}}, \bibinfo {author} {\bibfnamefont {G.}~\bibnamefont {Castaldini}}, \
  and\ \bibinfo {author} {\bibfnamefont {F.}~\bibnamefont {Braghin}},\
  }\bibfield  {title} {\enquote {\bibinfo {title} {Adiabatic edge-to-edge
  transformations in time-modulated elastic lattices and non-{H}ermitian
  shortcuts},}\ }\href {\doibase 10.1088/1367-2630/ac1ed4} {\bibfield
  {journal} {\bibinfo  {journal} {New J. Phys.}\ }\textbf {\bibinfo {volume}
  {23}} (\bibinfo {year} {2021}),\ 10.1088/1367-2630/ac1ed4}\BibitemShut
  {NoStop}%
\bibitem [{\citenamefont {Thouless}\ and\ \citenamefont
  {Kirkpatrick}(1981)}]{Thouless1981Conductivity}%
  \BibitemOpen
  \bibfield  {author} {\bibinfo {author} {\bibfnamefont {D.~J.}\ \bibnamefont
  {Thouless}}\ and\ \bibinfo {author} {\bibfnamefont {S.}~\bibnamefont
  {Kirkpatrick}},\ }\bibfield  {title} {\enquote {\bibinfo {title}
  {Conductivity of the disordered linear chain},}\ }\href {\doibase
  10.1088/0022-3719/14/3/007} {\bibfield  {journal} {\bibinfo  {journal}
  {Journal of Physics C: Solid State Physics}\ }\textbf {\bibinfo {volume}
  {14}},\ \bibinfo {pages} {235} (\bibinfo {year} {1981})}\BibitemShut
  {NoStop}%
\bibitem [{\citenamefont {Lin}\ and\ \citenamefont {Gong}(2024)}]{Lin2024Fate}%
  \BibitemOpen
  \bibfield  {author} {\bibinfo {author} {\bibfnamefont {Xiaoshui}\
  \bibnamefont {Lin}}\ and\ \bibinfo {author} {\bibfnamefont {Ming}\
  \bibnamefont {Gong}},\ }\bibfield  {title} {\enquote {\bibinfo {title} {Fate
  of localization in a coupled free chain and a disordered chain},}\ }\href
  {\doibase 10.1103/PhysRevA.109.033310} {\bibfield  {journal} {\bibinfo
  {journal} {Phys. Rev. A}\ }\textbf {\bibinfo {volume} {109}},\ \bibinfo
  {pages} {033310} (\bibinfo {year} {2024})}\BibitemShut {NoStop}%
\bibitem [{\citenamefont {Zhang}\ \emph {et~al.}(2018)\citenamefont {Zhang},
  \citenamefont {Kartashov}, \citenamefont {Zhang}, \citenamefont {Torner},\
  and\ \citenamefont {Skryabin}}]{ZhangY2018Resonant}%
  \BibitemOpen
  \bibfield  {author} {\bibinfo {author} {\bibfnamefont {Yiqi}\ \bibnamefont
  {Zhang}}, \bibinfo {author} {\bibfnamefont {Yaroslav~V.}\ \bibnamefont
  {Kartashov}}, \bibinfo {author} {\bibfnamefont {Yanpeng}\ \bibnamefont
  {Zhang}}, \bibinfo {author} {\bibfnamefont {Lluis}\ \bibnamefont {Torner}}, \
  and\ \bibinfo {author} {\bibfnamefont {Dmitry~V.}\ \bibnamefont {Skryabin}},\
  }\bibfield  {title} {\enquote {\bibinfo {title} {Resonant edge-state
  switching in polariton topological insulators},}\ }\href {\doibase
  https://doi.org/10.1002/lpor.201700348} {\bibfield  {journal} {\bibinfo
  {journal} {Laser \& Photonics Reviews}\ }\textbf {\bibinfo {volume} {12}},\
  \bibinfo {pages} {1700348} (\bibinfo {year} {2018})}\BibitemShut {NoStop}%
\bibitem [{\citenamefont {Privitera}\ \emph {et~al.}(2018)\citenamefont
  {Privitera}, \citenamefont {Russomanno}, \citenamefont {Citro},\ and\
  \citenamefont {Santoro}}]{Privitera2018Nonadiabatic}%
  \BibitemOpen
  \bibfield  {author} {\bibinfo {author} {\bibfnamefont {L.}~\bibnamefont
  {Privitera}}, \bibinfo {author} {\bibfnamefont {A.}~\bibnamefont
  {Russomanno}}, \bibinfo {author} {\bibfnamefont {R.}~\bibnamefont {Citro}}, \
  and\ \bibinfo {author} {\bibfnamefont {G.~E.}\ \bibnamefont {Santoro}},\
  }\bibfield  {title} {\enquote {\bibinfo {title} {Nonadiabatic breaking of
  topological pumping},}\ }\href {\doibase 10.1103/PhysRevLett.120.106601}
  {\bibfield  {journal} {\bibinfo  {journal} {Phys. Rev. Lett.}\ }\textbf
  {\bibinfo {volume} {120}},\ \bibinfo {pages} {106601} (\bibinfo {year}
  {2018})}\BibitemShut {NoStop}%
\bibitem [{\citenamefont {Lin}\ \emph {et~al.}(2022)\citenamefont {Lin},
  \citenamefont {Li}, \citenamefont {Xiao}, \citenamefont {Wang}, \citenamefont
  {Yi},\ and\ \citenamefont {Xue}}]{LinQ2022}%
  \BibitemOpen
  \bibfield  {author} {\bibinfo {author} {\bibfnamefont {Q.}~\bibnamefont
  {Lin}}, \bibinfo {author} {\bibfnamefont {T.}~\bibnamefont {Li}}, \bibinfo
  {author} {\bibfnamefont {L.}~\bibnamefont {Xiao}}, \bibinfo {author}
  {\bibfnamefont {K.}~\bibnamefont {Wang}}, \bibinfo {author} {\bibfnamefont
  {W.}~\bibnamefont {Yi}}, \ and\ \bibinfo {author} {\bibfnamefont
  {P.}~\bibnamefont {Xue}},\ }\bibfield  {title} {\enquote {\bibinfo {title}
  {Topological phase transitions and mobility edges in non-{Hermitian}
  quasicrystals},}\ }\href {\doibase 10.1103/PhysRevLett.129.113601} {\bibfield
   {journal} {\bibinfo  {journal} {Phys. Rev. Lett.}\ }\textbf {\bibinfo
  {volume} {129}},\ \bibinfo {pages} {113601} (\bibinfo {year}
  {2022})}\BibitemShut {NoStop}%
\bibitem [{\citenamefont {Wang}\ \emph {et~al.}(2018)\citenamefont {Wang},
  \citenamefont {Zhang},\ and\ \citenamefont {Song}}]{Wang2018Dynamical}%
  \BibitemOpen
  \bibfield  {author} {\bibinfo {author} {\bibfnamefont {R.}~\bibnamefont
  {Wang}}, \bibinfo {author} {\bibfnamefont {X.~Z.}\ \bibnamefont {Zhang}}, \
  and\ \bibinfo {author} {\bibfnamefont {Z.}~\bibnamefont {Song}},\ }\bibfield
  {title} {\enquote {\bibinfo {title} {Dynamical topological invariant for the
  non-{Hermitian} {R}ice-{M}ele model},}\ }\href {\doibase
  10.1103/PhysRevA.98.042120} {\bibfield  {journal} {\bibinfo  {journal} {Phys.
  Rev. A}\ }\textbf {\bibinfo {volume} {98}},\ \bibinfo {pages} {042120}
  (\bibinfo {year} {2018})}\BibitemShut {NoStop}%
\bibitem [{\citenamefont {Ivakhnenko}\ \emph {et~al.}(2023)\citenamefont
  {Ivakhnenko}, \citenamefont {Shevchenko},\ and\ \citenamefont
  {Nori}}]{Ivakhnenko2023Nonadiabatic}%
  \BibitemOpen
  \bibfield  {author} {\bibinfo {author} {\bibfnamefont {O.~V.}\ \bibnamefont
  {Ivakhnenko}}, \bibinfo {author} {\bibfnamefont {S.~N.}\ \bibnamefont
  {Shevchenko}}, \ and\ \bibinfo {author} {\bibfnamefont {F.}~\bibnamefont
  {Nori}},\ }\bibfield  {title} {\enquote {\bibinfo {title} {{Nonadiabatic
  Landau-Zener-St{\"u}ckelberg–Majorana} transitions, dynamics, and
  interference},}\ }\href {\doibase 10.1016/j.physrep.2022.10.002} {\bibfield
  {journal} {\bibinfo  {journal} {Physics Reports}\ }\textbf {\bibinfo {volume}
  {995}},\ \bibinfo {pages} {1--89} (\bibinfo {year} {2023})}\BibitemShut
  {NoStop}%
\end{thebibliography}

\providecommand{\noopsort}[1]{}\providecommand{\singleletter}[1]{#1}%

\end{document}